\documentclass[authorversion,nonacm]{acmart} 

\acmConference[Short Name]{Full Conference Name}{April 1, 2025}{Location}
\acmBooktitle{Proceedings of the Full Conference Name}

\usepackage{url}
\usepackage{tcolorbox}
\usepackage{wrapfig}
\usepackage{breakurl}
\usepackage{bold-extra}
\usepackage{lipsum}
\usepackage{amsfonts}
\usepackage{graphicx}
\usepackage{stfloats}
\usepackage{enumitem}
\usepackage{xspace}
\usepackage{comment}
\usepackage{caption}
\usepackage{epstopdf}
\usepackage{algorithmic}
\usepackage{algorithm2e}
\usepackage{tabularx}
\usepackage{setspace}
\usepackage{todonotes}
\usepackage{xcolor}
\usepackage{wrapfig}
\usepackage{array}
\usepackage{makecell}
\newcolumntype{P}[1]{>{\centering\arraybackslash}p{#1}}
\usepackage[flushleft]{threeparttable}
\usepackage{array,booktabs,makecell}

\usepackage{tabulary}
\usepackage[normalem]{ulem} 
\usepackage{multirow,array}
\usepackage{hyperref}   
\usepackage{pdflscape}  
\usepackage{colortbl}
\usepackage{subcaption}

\definecolor{light-gray}{gray}{0.96}
\definecolor{LightCyan}{rgb}{0.88,1,1}

\newcommand{\B}{\textsc{Binary Exponential Backoff}\xspace}
\newcommand{\LB}{\textsc{Log-Backoff}\xspace}
\newcommand{\LLB}{\textsc{LogLog-Backoff}\xspace}
\newcommand{\STB}{\textsc{Sawtooth-Backoff}\xspace}
\newcommand{\TSTB}{\textsc{Truncated Sawtooth-Backoff}\xspace}
\newcommand{\PB}{\textsc{Polynomial Backoff}\xspace}
\newcommand{\FB}{\textsc{Fixed Backoff}\xspace}

\newcommand{\beb}{BEB\xspace}
\newcommand{\llb}{LLB\xspace}
\newcommand{\lb}{LB\xspace}
\newcommand{\stb}{STB\xspace}
\newcommand{\tstb}{TSTB\xspace}
\newcommand{\fb}{FB\xspace}
\newcommand{\pb}{PB\xspace}

\newcommand{\thing}{participant\xspace}
\newcommand{\things}{participants\xspace}

\newcommand{\Things}{Participants\xspace}

\definecolor{light-gray}{gray}{0.96}

\renewcommand{\arraystretch}{1.4} 
\newcommand{\defn}[1]{\textbf{\emph{#1}}}
\newcommand{\poly}[1]{{\texttt{poly}({#1})}}
\newcommand{\whp}{w.h.p.\xspace}

\definecolor{darkbrown}{rgb}{0.45,0.2,0.35}
\definecolor{darkgreen}{rgb}{0.3,0.5,0.2}
\definecolor{ultramarine}{rgb}{0.07, 0.04, 0.76}

\newenvironment{revfontg}{\color{black}}

\newif\ifcomments
\commentsfalse

\AtBeginDocument{%
  \providecommand\BibTeX{{%
    \normalfont B\kern-0.5em{\scshape i\kern-0.25em b}\kern-0.8em\TeX}}}

\begin{document}


\title{A Survey on Adversarial Contention Resolution}

\author{Ioana Banicescu}
\email{ioana@cse.msstate.edu}
\orcid{}
\affiliation{%
  \institution{Mississippi State University}
  \streetaddress{Department of Computer Science and Engineering}
  \city{Mississippi State}
  \state{MS}
  \country{USA}
  \postcode{39762}
}

\author{Trisha Chakraborty}
\email{tc2006@msstate.edu}
\orcid{}
\affiliation{%
  \institution{Amazon Web Services}
  \streetaddress{}
  \city{}
  \state{MS}
  \country{USA}
  \postcode{}
}

\author{Seth Gilbert}
\email{gilbert@comp.nus.edu.sg}
\orcid{}
\affiliation{%
  \institution{National University of Singapore}
  \streetaddress{Department  of Computer Science, School of Computing}
  \country{Singapore}
  \postcode{119077}
}

\author{Maxwell Young}
\authornote{This work is supported by the NSF award CCF-2144410. Contact author: myoung@cse.msstate.edu.}
\email{young@cse.msstate.edu}
\orcid{}
\affiliation{%
  \institution{Mississippi State University}
  \streetaddress{Department of Computer Science and Engineering}
  \city{Mississippi State}
  \state{MS}
  \country{USA}
  \postcode{39762}
}

\renewcommand{\shortauthors}{Banicescu, Chakraborty, Gilbert, Young}

\begin{abstract}
Contention resolution addresses the challenge of coordinating access by multiple processes to a shared resource such as memory, disk storage, or a communication channel. Originally spurred by challenges in database systems and bus networks, contention resolution has endured as an important abstraction for resource sharing, despite decades of technological change. Here, we survey the literature on resolving worst-case contention, where the number of processes and the time at which each process may start seeking access to the resource is dictated by an adversary. We also  highlight the evolution of contention resolution, where new concerns---such as security, quality of service, and energy efficiency---are motivated by modern systems. These efforts have yielded insights into the limits of randomized and deterministic approaches, as well as the impact of different model assumptions such as global clock synchronization, knowledge of the number of processors, feedback from access attempts, and attacks on the availability of the shared resource. 
\end{abstract}



\maketitle

\clearpage


\section{Introduction}\label{sec:introduction}

\begin{itemize}[leftmargin=5.85cm]
\item[]{\it ``I saw two New Yorkers, complete strangers, sharing a cab. One guy took the tires and the radio; the other guy took the engine.''} \hspace{5pt} -- David Letterman
\end{itemize}

\medskip

How do we share a resource? This is a common problem, not only in large cities such as New York City, but also in a wide range of computer systems such as: wireless clients in WiFi networks~\cite{bianchi2000performance} and cellular networks~\cite{RamaiyanDo14}; processes in a shared-memory architecture~\cite{Ben-DavidB17}; nodes deciding on membership in a maximal independent set~\cite{daum2013maximal}; and concurrency control for databases~\cite{mitra:probabilistic,mitra:non-exlusive}. These are examples where system participants must share a resource, such as memory, disk storage, a data structure, or a wired or wireless communication channel.  \defn{Contention resolution} refers to this challenge of coordinating access to a common resource. 

The origins of contention resolution lie with initial research on  databases and old Ethernet  networks~\cite{kurose:computer}. However, the problem has grown in importance with the adoption of wireless communication, where the resource under contention is a broadcast channel. This began with the ALOHA protocol~\cite{abramson1985development}, which initiated an  effort in the research community to understand the performance of ALOHA, particularly when packet arrival times are dictated by a stochastic process (for examples, see \cite{ALOHAnet,tsybakov1979ergodicity,ferguson1975control,fayolle1977stability,abramson1977throughput,rosenkrantz1983instability,kelly1987number,kelly:decentralized}). A foundational algorithm from this time period is randomized \defn{binary exponential backoff (\beb)} \cite{MetcalfeBo76}, which continues to play an important role in wireless communication. Subsequently, a number of variants were designed and analyzed, such as polynomial backoff \cite{haastad1996analysis}. The core ideas used by these older algorithms of multiplicatively increasing and decreasing the sending probability (i.e., ``backing on'' and ``backing off'')  remain key building blocks for many recent results.

While models with stochastic arrivals received significant attention from the theory community, empirical studies emerged  to suggest that traffic often exhibits bursty behavior~\cite{martin2023self,Teymori2005,yu:study}, where a large batch of packets arrive simultaneously. Thus, researchers  considered a setting where a batch of packets arrive together and must be successfully transmitted on the shared channel before the next batch arrives, and so on. While this burstiness is still somewhat well-behaved---since batches never overlap---it posed new challenges over the stochastic setting. A surprising result by Bender et al. \cite{bender:adversarial} demonstrated that \beb performs sub-optimally, requiring $\Omega(n\log n)$ slots to complete a single batch of $n$ packets. This spurred interest in better  algorithms for batched arrivals, and we discuss these results in Section~\ref{sec:backoff-algorithms}. 

Over the past decade and a half, attention has turned to the case where the arrival times of packets are far-less constrained. One common approach has been to view this problem through the lens of adversarial queuing theory~\cite{borodin2001adversarial}, where results are parameterized by traffic characteristics such as the arrival rate and the degree of burstiness. Another major branch in the literature addresses worst-case arrival times. Arguably, deterministic approaches have been influenced by the pioneering result of Koml\'{o}s and Greenberg \cite{komlos:asymptotically}, leading to new solutions for worst-case arrival times. A prime example is the work by De Marco and Kowalski \cite{doi:10.1137/140982763}, and there has been a long line of results in this area (e.g., \cite{DBLP:conf/icdcs/MarcoKS19,Chlebus:2016:SWM:2882263.2882514,DEMARCO20171,de2013contention}). In parallel, many different randomized approaches have been explored. A recent influential paper by Chang et al. \cite{ChangJP19} shows how the classic ingredients of backing off/on can yield an elegant solution to worst-case arrivals. The literature in these areas is explored in Sections~\ref{sec:aqt-results}, \ref{sec:deterministic} and \ref{sec:random-stream}.

More recently, the problem has evolved further to capture the challenges motivated by modern systems. For example, energy efficiency is important to many ad-hoc and mobile network settings, where devices are typically battery-powered \cite{SINHA201714,kenny:performance,polastre:telos}.  Another concern is malicious interference, which amounts to a denial-of-service attack~\cite{pirayesh2022jamming,arjoune2020smart}. Early work in this area by Awerbuch et al. \cite{awerbuch:jamming}, leading to subsequent results by  Richa et al. \cite{richa:jamming2,DBLP:journals/dc/RichaSSZ13,richa:jamming3,DBLP:journals/ton/RichaSSZ13,richa:jamming4} and Ogierman~et al.~\cite{ogierman:competitive,ogierman2018sade}, showed how such interference can be tolerated, and attack-resistant contention resolution has since become an active area of research. These issues and others discussed in Sections~ \ref{sec:random-stream} and \ref{sec:jamming}.

\smallskip

\subsubsection*{\bf Scope and Goals.} In this survey, we focus primarily on contention resolution in the presence of an adversary that (1) controls when \things can begin executing their contention-resolution protocol, and (2) may deny access to the shared resource for periods of time; we refer to this as \defn{adversarial contention resolution}. This stands in contrast to the setting where arrival times are governed by a stochastic process, which is an area of research that has been previously surveyed by Chlebus~\cite{chlebus2001randomized}. 

{\revfontg 
What motivates our focus on adversarial contention resolution? From a theoretical standpoint, it is interesting to explore what is possible in such a challenging setting, but there are also practical considerations for considering an adversarial context. Network traffic does not always align with stochastic models; specifically, bursty traffic fails to be captured by well-behaved (e.g., Poisson- or Bernoulli-distributed) arrival times. Additionally, the mobility of many modern computing devices, along with connectivity challenges in wireless systems, can lead to situations where the contention for a shared resource changes rapidly over time.}   Given this motivation, our overall goal in this survey paper is to summarize the models and performance metrics in the literature on adversarial contention resolution, along with discussing their corresponding theoretical results.

\subsection{Definitions, Terminology,  Models, Metrics, and Algorithm Design}\label{sec:prelim}

Contention resolution has gone by other names in the literature, such as  $k$-selection~\cite{FernandezAntaMoMu13}, all-broadcast~\cite{chlebus2001randomized}, and conflict resolution~\cite{komlos:asymptotically}. To provide a unified presentation, we review  terminology that is common to the literature, elaborate on typical model assumptions and performance metrics, and highlight those aspects used to organize the literature.

\subsubsection{Multiple Access Channel.} Throughout, we consider a time-slotted system with {\boldmath{$n$}} $\geq 2$ system {\defn{\things}}, each seeking access to a shared resource. In each \defn{slot}, a \thing may choose to access the shared resource and, if it is the only \thing to do so, this access attempt is \defn{successful}. Else, if two or more \things make an attempt in the same slot, the result is a \defn{collision} and no \thing succeeds in this slot. Finally, if no \thing makes an access attempt, then the corresponding slot is \defn{empty}.

In practice, the resource itself may be local to a single machine; for example, corresponding to the critical section of a program, or a common disk shared by multiple processes. This extends to networked settings where \things access shared memory or separate machines via remote procedure calls. 

In the literature, the resource is frequently treated as a shared communication channel, and this is the convention we use throughout this survey. Such a channel is typically referred to as a \defn{multiple access channel}, which is a popular model under which several fundamental problems have been examined, such as mutual exclusion~\cite{bienkowski2010dynamic,czyzowicz2009consensus}, leader election~\cite{chang:exponential-jacm,chang:exponential,nakano2002}, consensus~\cite{czyzowicz2009consensus}, routing~\cite{chlebus2019energy}, broadcast \cite{Marco08,Marco10,ChlebusGGPR00,alon1991lower,ClementiMS03,DBLP:conf/podc/Kowalski05,chrobak2002fast,piotr:explicit,chlebus2005almost,kowalski2005time}, and well-known random access protocols~\cite{kurose:computer}. Here, for any given slot, accessing the resource corresponds to a \thing \defn{listening} (equivalently, \defn{receiving}) or \defn{sending} (equivalently, \defn{transmitting}) a \defn{packet} on the multiple access channel. Unless stated otherwise, a packet can be transmitted within a single slot. When a packet is sent in a slot, we typically say that this packet succeeds if no other packet is sent in the same slot; otherwise, the packet fails.

For ease of exposition, we will often slightly abuse terminology by speaking of packets themselves taking action (e.g., sending themselves on the channel, or listening on the channel). This ``\defn{packet-centric}'' presentation allows us to avoid referring to the corresponding  ``\things''/``devices''/``nodes'' that actually perform these actions.


\subsubsection{Models Assumptions.}\label{sec:model-assumptions} We discuss model assumptions that play an important role in establishing upper and lower bounds with respect to the metrics discussed above.  \smallskip

\noindent{\it Packet Arrivals.} An important aspect is the timing by which new packets are generated (i.e., arrive) in the system. A simple case is where each \thing has a single packet, and all \things start executing a contention-resolution algorithm at the same point in time. Equivalently, from a packet-centric view, all packets arrive simultaneously as a single \defn{batch}; this is often referred to as the \defn{static case} in the literature (see~\cite{DBLP:conf/podc/Kowalski05}). 

In contrast, in the \defn{dynamic} case, packets arrive over time; this arrival model is also referred to in the literature as \defn{asynchronous} (see ~\cite{DBLP:conf/podc/Kowalski05}). The literature adopts the terminology of \defn{arrival time} or \defn{activation time} to indicate when the corresponding \thing begins executing a contention-resolution protocol in order to send this packet. These times may be governed by a stochastic process (see the survey by Chlebus~\cite{chlebus2001randomized}), or they may be scheduled by an adversary; in this survey, we focus on the latter setting.\medskip

\noindent{\it Channel Characteristics.} When a \thing sends a packet, the multiple access channel provides feedback. The most basic feedback---assumed in all the literature that we survey---occurs when, upon sending its packet, a \thing learns whether or not its transmission was successful.

Another type of feedback concerns whether \things can detect collisions. Almost all of the literature on contention resolution assumes one of the following two types of feedback regarding collisions. The first is \defn{collision detection (CD)}: in any slot,  each  listening \thing  can differentiate between the case where the number of \things transmitting is $0$ (i.e., an empty slot) or greater than $1$ (i.e., a collision). The second is \defn{no collision detection (no-CD)}: in any slot, a listening device cannot differentiate between the slot being empty versus the slot containing a collision.  

In many works, \things that listen to the channel also learn whether or not a slot contains a successful transmission. In such a setting with CD, the channel provides {\it ternary feedback} (sometimes referred to as ``trinary feedback'') indicating whether the slot is empty, contains a successful transmission, or contains a collision. Similarly, in this setting, no-CD provides {\it binary feedback}.\footnote{Other variants of CD appear in the wider literature involving multiple access channels (see ~\cite{chang:exponential,chang:exponential-jacm}). Additionally,  more exotic channel models have been considered; for example, De Marco and Kowalski~\cite{DBLP:conf/icdcs/MarcoK10} and Tsybakov \cite{tsybakov:multiplicity} examine the case where channel feedback yields the number of senders.} 



Finally, some papers consider the case where more than one channel is available to \things; this is typically referred to as a \defn{multi-channel setting}. Here, typically \things may select any one (and typically only one) channel to send on or listen to in any give slot. {\revfontg A result involving a single channel may extend to the multi-channel setting without altering the core finding; however, there {\it are} cases where the use of multiple channels can admit significantly stronger results. Typically, this occurs when multiple channels are leveraged to achieve better latency, or robustness to malicious interference. In practice, multiple channels are available on several  communication standards, such as IEEE 802.11 (WiFi), Bluetooth, and cellular standards (e.g., 4G and 5G LTE). Therefore, there is real-world motivation to push for results that make good use of more than one channel. However, in comparison to the single-channel setting, there are fewer results in the contention-resolution literature involving multiple channels.}

\medskip

\noindent{\it Network Topology.} A \defn{single-hop network} is one where each \thing can communicate with every other \thing directly. In contrast, in a \defn{multi-hop network}, a communication may need to pass between many \things until it reaches its intended recipient. 

For the contention-resolution problem, the type of topology may impact channel feedback. With CD in a single-hop network, all \things can detect the same channel feedback in any fixed slot. {\revfontg However, in a multi-hop network, it is common for different sets of \things to witness different channel feedback and, therefore, potentially update their respective states differently.  Given this challenge, it is not surprising that the majority of work on contention resolution has focused the single-hop setting.}  

\medskip

\noindent{\it Global Knowledge.} What information do all packets have about the system {\it a priori}? Having all packets agree on certain system parameters can significantly ease the challenge of contention resolution.

\begin{itemize}[leftmargin=18pt]
\item{\it System-Size Knowledge.} The exact number of the \things may be known {\it a priori}. Alternatively, an upper bound may be common knowledge. In some problem formulations (see Section~\ref{sec:deterministic}), only a subset of {\boldmath{$k\leq n$}} \things have a packet to send, where $n$ acts as an upper bound that is known or tightly-estimated {\it a priori}. \smallskip

\item {\it Unique Identifiers.} Given {\boldmath{$n$}} \things, each may have a unique identifier (\defn{ID}) taken from a set of $\Theta(n)$ possible IDs.  This ID may be used by an algorithm to determine when the corresponding station accesses the channel.\smallskip

\item{\it Global Clock.} In the static case, all \things obtain their respective packet in the same slot $s$. This offers a common point for synchronization; that is, all \things may consider $s$ to have index $0$. Thus, for the static case, there is the implicit assumption of a global clock. For dynamic arrivals, packets may be generated at \things at different times. In this case, if all \things agree on the index of the current slot, then they are said to have access to a global clock; otherwise, they are often said to have only a \defn{local clock}.
\end{itemize}


\subsubsection{Metrics.}\label{sec:metrics}  In the static case, a natural performance measure is the number of slots until a set of packets succeeds; this number of slots is referred to as the \defn{makespan}. We are typically interested in the set of all packets, but any deviations from this will be made clear from the context.  For the static case, the ratio of the number of successful packets to the makespan is known as \defn{throughput}, and this is another common notion of performance; note that throughput is a value between $0$ and $1$. 

For the dynamic case, we define a slot to be \defn{active}, if at least one packet is present in the system during that slot. The above metrics remain valid when measured over active slots.  However, at intermediate points in time, throughput may offer an overly-pessimistic measure of progress. For example, if a batch of $n$ packets arrive at the same time and starts executing \beb, the throughput will be $0$ for $\Omega(n/\log n)$ slots, despite the fact that the algorithm is progressing towards having many successful packet transmissions, which occurs after $\Theta(n)$ slots are executed.

To provide a more meaningful metric, Bender et al.~\cite{bender2020contention} propose \defn{implicit throughput}, which at time $t$ is defined as $N_t/A_t$, where $N_t$ is the total number of packets that arrive at or before time~$t$, and $A_t$ is the total number of active slots at or before time~$t$. Returning to the example with a batch of $n$ packets, we note that the implicit throughput is large---specifically, $n$---in the first slot, and is $\Omega(\log n)$ over the first $\Omega(n/\log n)$ slots. The implicit throughput will continue to decrease, reaching  {\revfontg $\Theta(1)$}. Then, ultimately, the implicit throughput will be $\Theta(1/\log n)$ once all packets have succeeded, which matches the standard measure of throughput for \beb, as mentioned in Section \ref{sec:introduction}. An algorithm that guarantees an implicit throughout of $\Omega(1)$ over all slots performs well, since the number of active slots so far is never asymptotically larger than the number of packets that have arrived. Furthermore, as exemplified above, at points in time where the standard notion of throughput is meaningful, the two metrics are equal.

Continuing with the dynamic case, another metric is the number of slots it takes for a \thing to successfully transmit its packet starting from its activation time. We are interested in the maximum such number of slots over all \things; we refer to this as (maximum)  \defn{latency} . Here, we can extend throughput to be the number of packets over the latency. 

In much of the early literature on the dynamic case, packets are generated according to a stochastic process, and efficiency is tied to the \defn{packet arrival rate}; that is, the number of  packets that are newly generated (\defn{arrive}) in each slot and held by (some subset) of \things. For any \thing, its unsent packets are stored in a queue for eventual transmission on the multiple access channel, and this number of unsent packets is referred to as the \defn{queue size}. Given a packet arrival rate, the goal is to guarantee that every queue has bounded size. The higher the rate that can be tolerated, the more efficiently the algorithm allows the network to digest new packets.  In Section~\ref{sec:aqt-results}, we expand on this notion, which is generally referred to as \defn{stability}.

Given the popularity of small, battery-powered wireless devices in modern networks, many contemporary results focus on energy efficiency. The energy consumed by a \thing is typically measured as the number of slots that the \thing accesses the channel, since sending and listening often dominate the operational costs of such devices~\cite{SINHA201714,feeney2001investigating,polastre:telos,wang2006realistic}; we refer to this as the \defn{number of channel accesses}.

Robustness to malicious interference---often referred to as \defn{jamming}---is becoming increasingly important, with several well-publicized  attacks in recent years~\cite{nicholl,FCC:humphreys,marriott,jamming-church}. Contention-resolution algorithms that rely on channel feedback are especially vulnerable to adversarial noise. We refer to this aspect as \defn{jamming resistance} and, for each relevant result, we describe the type of attack that can be tolerated.


\subsubsection{Algorithm Design.}\label{s:alg-design} There is some common terminology used in the literature to characterize different classes of contention-resolution algorithms. An obvious division is with respect to deterministic versus randomized algorithms, where the former is characterized by the absence of random bits being used in the execution. Deterministic solutions are appealing from a practical standpoint in that their guarantees hold even when the number of \things and/or packets is small, admitting no error probability. In trade, they tend to provide sub-optimal asymptotic results in many settings, compared to randomized solutions, as we shall see.


An \defn{adaptive} algorithm knows the full history of channel feedback and can use this information to inform future actions. Messages other than the sending of the packet itself are allowed; for example, such messages might pertain to estimates of the system size or coordination information. Sometimes this content is referred to as \defn{control bits}~\cite{hradovich2020contention}, and they are often used to facilitate coordination (see~\cite{chlebus:adversarial}). In contrast, the only feedback that might alter the execution of a \defn{non-adaptive} algorithm is a \thing learning that its packet has been successfully  transmitted, after which the \thing terminates~\cite{doi:10.1137/140982763}. There is a less-common notion of \defn{strong non-adaptiveness}, where the execution of a \thing is not even altered by the outcome of a transmission (again, see~\cite{doi:10.1137/140982763}).  \smallskip

\begin{figure}[t!]
\captionsetup[subfigure]{labelformat=empty}
\begin{subfigure}{1.0\textwidth}
\centering
\includegraphics[scale=0.4]{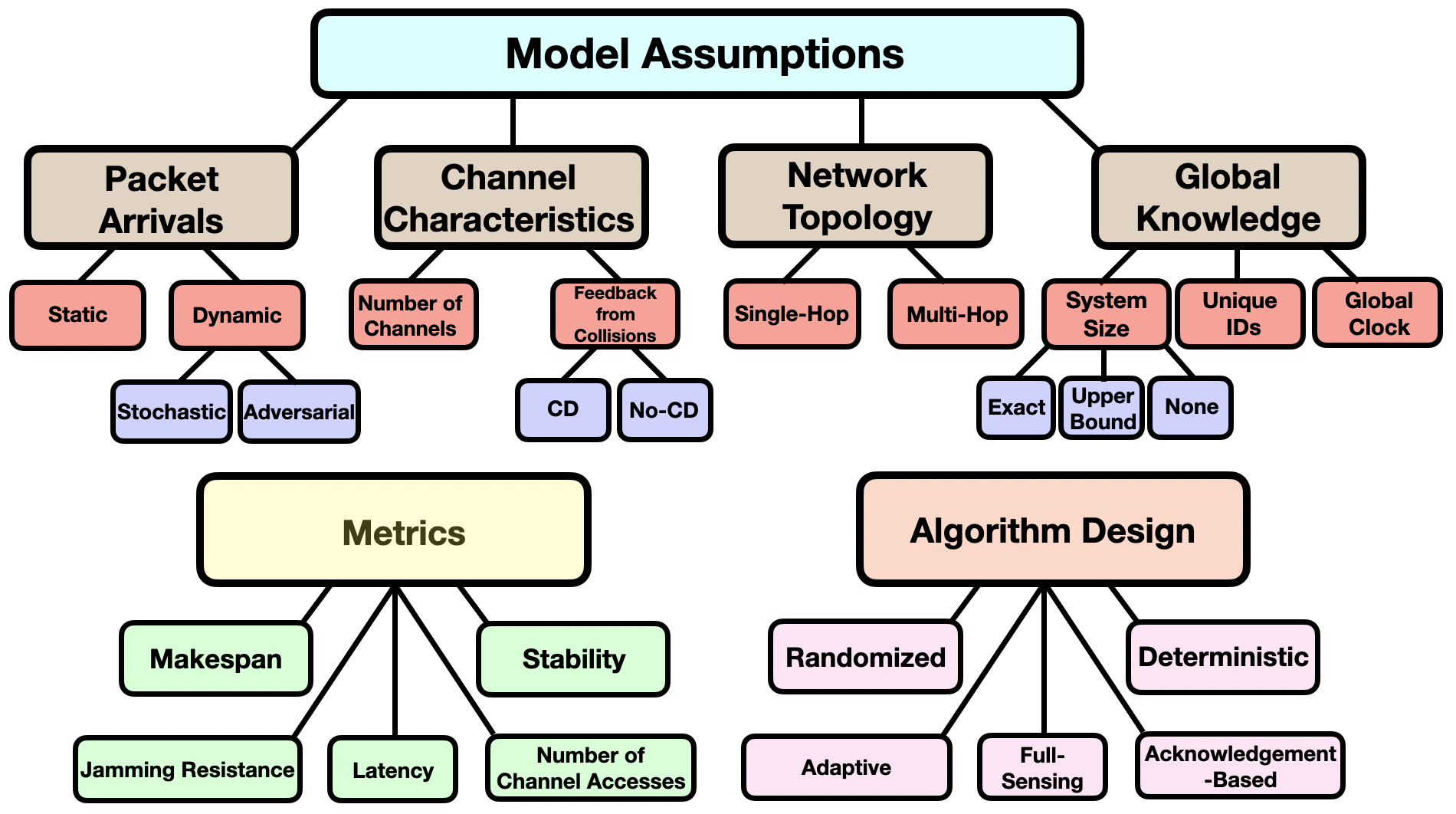}
\end{subfigure}
\caption{An illustration of features used to structure our survey of the literature. Larger boxes correspond to general features, while smaller boxes indicate those that are more fine-grained.}
\vspace{-5pt}
\label{fig:structure}
\end{figure}

Finer distinctions are sometimes used.  A \defn{full-sensing} algorithm has the same characteristics as an adaptive algorithm, but no control bits are allowed~\cite{chlebus2006adversarial,hradovich2020contention}. Finally, an algorithm where packets can use acknowledgments (implicitly received after succeeding) to terminate is  said to be \defn{acknowledgement-based}~\cite{DBLP:conf/icdcs/MarcoKS19}; otherwise, if packets continue executing after their respective successes, the algorithm is \defn{non-acknowledgement-based}.\footnote{This terminology varies slightly in the literature. In the stochastic dynamic model, an \defn{acknowledgement-based} algorithm depends on the packet's own transmission history~\cite{Goldberg-notes-2000}. In \defn{adversarial queueing theory (AQT)} (see Section~\ref{sec:aqt-results}), it means a packet is sent according to a schedule based on its ID and the current round~\cite{chlebus2006adversarial,DBLP:conf/icdcs/MarcoKS19,
hradovich2020contention}.  The latter definition incorporates a dependency on IDs, since much of the work in the acknowledgement-based setting involves  deterministic algorithms, which require unique IDs (see Section~\ref{sec:deterministic} for more discussion). }
  
These classifications can be useful, since they indicate the strength of the underlying assumptions being made. For example, an algorithm that is randomized and full-sensing makes stronger assumptions than an algorithm that is deterministic and acknowledgement-based. A device running the former algorithm must have a pseudorandom number generator and be able to receive feedback (beyond learning of  its own success) on the channel. In contrast, the latter algorithm can function even when channel sensing is unavailable---either due to the nature of the channel or a \thing's hardware constraints---and the resulting protocol may be less complex in its implementation by eschewing randomness and knowledge of transmission history. 

However, we note that this classification can be complicated to apply in certain cases; for example, in a multi-channel setting,  a device may not have the full history of feedback on {\it all} channels. Therefore, in situations where the classification is clear, we present this information; otherwise, we omit it.


\subsection{Survey Organization}

We use the attributes discussed above and illustrated in Figure~\ref{fig:structure} to structure our review of the literature.  In Section~\ref{sec:backoff-algorithms}, we begin with contention-resolution algorithms for the static case. This provides an introduction to the important metric of makespan, as well as the use of randomization in designing contention-resolution algorithms. Next, in Section~\ref{sec:adversarial-dynamic}, we move to the dynamic model of packet arrivals, focusing on the case when arrival times are dictated by an adversary; here, latency is a key metric. Finally, in Section~\ref{sec:jamming}, we address additional adversarial behavior in the form of jamming. 

Each section has the following  associated components: 
\begin{enumerate}[leftmargin=12pt]
\item A specific result that we feel provides some intuition for an important aspect of the results surveyed. This helps us to highlight (i) those challenges posed by the  contention-resolution problem being addressed, and (ii) one way in which these challenges can be overcome, which in turn, allows us to contrast against some other surveyed solutions. 

\item A summary portion, where we highlight the main themes of the literature surveyed.

\item A table that offers a synopsis of the results in each section by organizing the results along common, pertinent attributes from Section~\ref{sec:prelim}. Due to space constraints, these tables can be found in our electronic supplement. 
\end{enumerate}


\section{The Static Case}\label{sec:backoff-algorithms}

We begin our survey with the static case for packet arrivals. This motivates a well-known class of randomized, non-adaptive \defn{backoff} algorithms, with origins in decades-old networking protocols; we explore this topic in Section~\ref{sec:backoff}. Additionally, the static case  also allows us to introduce deterministic algorithms, which are newer additions to the literature; we review these results in Section~\ref{sec:deterministic-static}. Finally, we discuss the closely-related wakeup problem in Section \ref{sec:static-wakeup}.

\subsection{Backoff Algorithms}\label{sec:backoff}

\begin{figure}[t]
\centering
\includegraphics[scale=0.4]{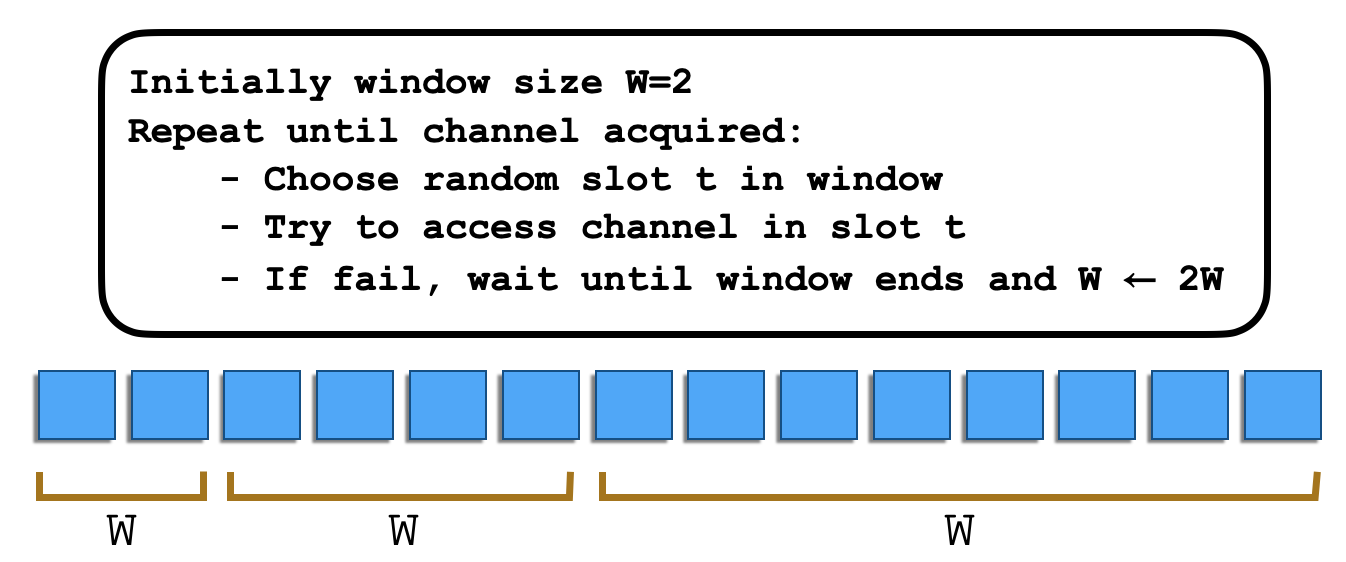}
\vspace{-5pt}\caption{Pseudocode for \beb and an illustration of the first few contention windows. Time moves from left to right.}\label{fig:beb}
\end{figure}

 Arguably, the most well-known backoff algorithm is binary exponential backoff (\beb), which was introduced by Metcalf and Boggs~\cite{MetcalfeBo76}, and briefly discussed in Section~\ref{sec:introduction}. \beb proceeds in disjoint \defn{contention windows (CWs)} consisting of consecutive slots. In each CW, a \thing with a packet selects a slot uniformly at random to attempt a transmission. If successful, another packet may be processed in the same fashion, if one exists. Else, transmission is attempted in the next CW, which is double the size of the current CW. This is illustrated in Figure~\ref{fig:beb}, along with pseudocode.

While originally intended for use with Ethernet over bus networks~\cite{stuck1984introduction}, \beb has since found application in a range of domains such as wireless communication~\cite{zhou2019singletons,sun:backoff,6859627,saher:log}, transactional memory~\cite{herelihy:transactional,scherer:advanced}, concurrent memory access~\cite{Ben-DavidB17,80120,mellor-crummey:algorithms}, and congestion control~\cite{mondal:removing}.\footnote{A ubiquitous example is WiFi, where many of the standards use BEB as a core mechanism for governing how devices gain access to a wireless channel. In these standards, the initial window size is typically some constant larger than $1$, and the maximum window size is limited to a value just over $1000$.}  Despite this widespread use, the makespan for \beb is sub-optimal (see Bender et al.~\cite{bender:adversarial}); specifically, for a batch of $n$ packets that begin executing \beb at the same time, the makespan is $\Theta(n \log n)$.  To understand why, we need to become familiar with the notion of contention.


\subsubsection*{\bf Developing Intuition.} In order to  better understand why \beb performs poorly, we consider a class of \defn{Bernoulli-backoff} algorithms (see \cite{bender:adversarial}), where each packet sends with probability $1/w$ in each slot of a window of size $w$. Note that this differs subtly from the approaches discussed in Section~\ref{sec:general-monotonic-backoff}, where  a single slot is chosen uniformly at random in which to send. For Bernoulli-backoff algorithms it is easy to analyze the useful notion of \defn{contention}~\cite{richa:jamming4,ChangJP19,BenderFGY19,DBLP:conf/spaa/AgrawalBFGY20,bender:how}. Let {\boldmath{$p_i(t)$}} denote the probability that packet $i$ attempts to broadcast in slot $t$. We define the contention in slot $t$ to be {\boldmath{$C(t)$}} $= \sum_i p_i(t)$; that is,  the sum of the sending probabilities for all packets in the system at time $t$.


Contention characterizes the probability that some packet is successful in slot $t$; denote this quantity by {\boldmath{$p_{\mbox{\tiny suc}}(t)$}}. In particular, if $p_i(t) \leq 1/2$, which is true if we set the initial window to be size at least $2$, then the following relationship holds:
\begin{align}
\frac{ C(t)}{e^{2C(t)}} &\leq p_{\mbox{\tiny suc}}(t) \leq \frac{2C(t)}{e^{C(t)}}.\label{eq:success-prob}
\end{align}
\vspace{0pt}

While we use Equation~\ref{eq:success-prob} here to gain intuition about the static case, these bounds on contention will also be useful later on when discussing the dynamic case (Section~\ref{sec:adversarial-dynamic}). Like the temperature of the porridge in the Goldilocks fable, 
the contention can be ``too high'' or ``too low'' for packets to succeed. For example, consider the Bernoulli equivalent of \beb. Initially, the contention is high. In particular, $C(t) = \Omega(\ln n)$ until a window of size $\Theta(n/\ln n)$, implying  that $p_{\mbox{\tiny suc}}(t)$ is polynomially small (or smaller) in $n$ during this portion of the execution. 

When the window size reaches $\Theta(n)$, $C(t)=\Theta(1)$ and so $p_{\mbox{\tiny suc}}(t) = \Theta(1)$. At this point, the contention is ``just right'' and many packets succeed.  Unfortunately, since the window size grows aggressively under \beb, the contention rapidly decreases, and so, any remaining packets require many slots before they succeed. This behavior is what leads to suboptimal makespan, and it begs the question: is there a backoff algorithm that yields the asymptotically optimal makespan of $O(n)$ or, equivalently, $\Omega(1)$ throughput? As a first step towards answering this question, we consider  backoff algorithms where the growth of contention windows differs from \beb.


\begin{figure}[t]
\begin{tcolorbox}[standard jigsaw, opacityback=0]

\noindent{}\hspace{-3pt}{\bf A Generic Backoff Algorithm}\medskip


\noindent\hspace{-3pt}$i$: index of current window for $i\geq 0$.

\noindent\hspace{-3pt}$w_i$: size of window $i$.\vspace{0pt}

\noindent\hspace{-3pt}$f(i)$: window-scaling function.\medskip

\noindent{}\hspace{-3pt}In current window $i$, a station with a packet to transmit does the following until successful:\vspace{-0pt}	 
	
          \begin{itemize}[leftmargin=3mm]
			
                 \item Attempt to transmit in a slot chosen uniformly at random from $w_i$.\vspace{-0pt}
			
                  \item If the transmission failed, then wait until the end of the window and $w_{i+1} \leftarrow  (1+ f(i))w_i$.
       
          \end{itemize}
\end{tcolorbox}
\vspace{-5pt}
\caption{Generic backoff with monotically-increasing windows. For  \lb, \llb, \beb, and \textsc{PB}, the initial window size is a constant, and $f(i)=1/\lg w_i, 1/\lg\lg w_i$, $1$, and $(1+1/i)^c - 1$, respectively. For \fb, $f(i) = 0$ and the initial window size is $\Theta(n)$.}\label{fig:pseusocode1}
 \end{figure}


\subsubsection*{\bf General Monotonic Backoff}\label{sec:general-monotonic-backoff}

The makespan of a backoff algorithm depends on the amount by which successive CWs increase in size; for \beb, this is simply a factor of $2$. Several results examine different values~\cite{Ni:survey,1543687,sun:backoff,6859627,saher:log}, presenting a mixture of empirical and analytical results when the network is \defn{saturated}; that is, when each \thing always has a packet to send~\cite{bianchi1998ieee}. In contrast, the static model of  arrivals captures a type of bursty traffic, which turns out to be more challenging. In this setting, Bender et al.~\cite{bender:adversarial} derive makespan results  for \beb and several closely-related variants:   \LB ~(\lb), \LLB~(\llb), and \FB~(\fb). These variants operate similarly to \beb: when the current CW of size $w$ is finished, the next CW has size $(1+ f(w))w$, where $f(w)=1/\lg(w), 1/\lg\lg(w),$ $1$, and $0$, for \lb, \llb, \beb, and \fb, respectively.\footnote{We use $\lg(x)$ to denote the logarithm base 2 of $x$.} In each CW, a \thing randomly selects a single slot in which to send, and other than learning of success or failure, no channel feedback is used by the algorithms; all of these algorithms are acknowledgement-based.

The pseudocode for these backoff algorithms is  presented in Figure~\ref{fig:pseusocode1}.  To the best of our knowledge, the initial CW sizes for \lb, \llb, and \fb are not specified exactly in the literature; $\Theta(1)$ will suffice for $\lb$ and $\llb$. A conspicuous requirement of \fb is that $n$ must be known to within a constant multiplicative factor in order to set the CW size. A number of prior works show how to get an estimate of $n$~\cite{DBLP:journals/jacm/GreenbergFL87,bender:contention,cali:dynamic,cali:design,bianchi:kalman}.

In terms of performance, for a batch of $n$ packets, the makespan for \beb, \lb, \fb, and \llb is $\Theta(n\log n)$, $\Theta\left(\frac{n\log n}{\log\log n}\right)$, $\Theta\left(n\log\log n\right)$, and $\Theta\left(\frac{n\log\log n}{\log\log\log n}\right)$, respectively~\cite{bender:adversarial,zhou2019singletons}. Bender et al. also analyze {\it polynomial} backoff (\textsc{PB}), where the $i^{\mbox{\tiny th}}$ window has size $\Theta(i^c)$ for some constant $c>1$. The window scaling function is slightly different here; ignoring constants, we solve for $f(i)$ in $(i+1)^c = (1+f(i)) i^c$, implies that $f(i) = ((1+i)^c / i^c) - 1  = ((i+1)/i)^c  - 1$. For \textsc{PB}, Bender et al. show that the makespan is  $\Theta(n/\log n)^{1+(1/c)}$.

Interestingly, all of these algorithms suffer from a throughput that tends to zero. This observation is striking for a couple reasons. First, it illustrates the challenges raised by the static case. In particular, the throughput for \pb when $c=2$, known as \defn{quadratic backoff}, is shown by Sun and Dai~\cite{sun:backoff} to be competitive with \beb under a saturated network. In contrast,  under static arrivals, quadratic backoff has asymptotically inferior makespan.

Second, from a theoretical perspective, does a backoff algorithm exist that gives optimal throughput under the static case? Bender et al.~\cite{bender:adversarial} show that, over the class of monotonic windowed backoff algorithms, \llb achieves the asymptotically-best makespan
~\cite{bender:adversarial}. In other words, a non-monotonic approach is required to achieve $\Theta(n)$ makespan or, equivalently, $\Theta(1)$ throughput. This question of asymptotic optimality is what we explore next.


\subsubsection*{\bf Understanding Performance.} To develop intuition for monotonic backoff algorithms, we start by asking: why does  \lb fare better than \beb? Considering the Bernoulli versions, initially the high contention prevents packets from succeeding. Indeed, the high contention persists for longer under \lb given that the window size grows far more slowly compared to \beb. However, once (near) constant contention is achieved, packets succeed. Critically, packets stay within the Goldilocks zone for longer since, again, the window size increases more slowly relative to \beb. Finally, contention drops off more gradually, allowing any straggling packets to succeed within fewer slots than under \beb.

The same qualitative behavior is true of \llb, which grows its window size faster than \lb, but more slowly than \beb. This allows \llb to strike the asymptotically optimal balance between the number of wasted slots spent in either high-or-low-contention windows, and those slots with constant contention.

Why doesn't \fb perform better? After all, it avoids the high-contention slots by skipping directly to a window of size $\Theta(n)$ where the probability of success is constant.  Here, again, contention tells the story. While initially many packets succeed under \fb, the contention quickly decreases, and this inflicts an $O(\log\log n)$-factor over the optimal makespan.

{\revfontg Ultimately, for the static case, the above algorithms fall short due to their inability to provide constant contention over $cn$ slots for some sufficiently large constant $c$.  Next, we examine algorithms that do satisfy this property.}


\subsubsection*{\bf Achieving Asymptotically Optimal Makespan}\label{sec:backoff-backon} 

Based on our observations in Section~\ref{sec:backoff-algorithms}, any algorithm that hopes to achieve $O(n)$ makespan must incorporate a mechanism for providing constant contention over $\Omega(n)$ slots. \STB (\stb)~\cite{GreenbergL85,Gereb-GrausT92} accomplishes this; it is  non-monotonic algorithm that executes over a doubly-nested loop. The outer loop sets the current window size $w$ to be double the one used in the preceding outer loop; this is like BEB. Additionally, for each such window, the inner loop executes over $\lg w$ windows of decreasing size: $w, w/2, w/4, ..., 1$. For each such window, a slot is chosen uniformly at random for the packet to transmit; this is the ``backon'' component of STB.
When the outer loop achieves a window of size $cn$ for a sufficiently large constant $c>0$, with high probability (i.e., probability at least $1-O(1/\poly{n})$, an abbreviated as \defn{\whp}), all packets will succeed in the corresponding inner loop. Intuitively, this is because the backon component maintains constant contention in each slot of the inner-loop windows (until all packets succeed). For a single batch of $n$ packets, the makespan of \stb is $O(n)$~\cite{Gereb-GrausT92,GreenbergL85,bender:adversarial,zhou2019singletons}.

A variant called \textsc{Truncated Sawtooth-Backoff} (\tstb)~\cite{bender:contention} executes identically to STB except that, for some constant $c>0$, the inner loop  executes over $\lg(c\lg w)$ windows of decreasing size: $w, w/2, w/4, ..., \max\{\lfloor w/(c\lg w) \rfloor, 1\}$. Another asymptotically-optimal result is given by Anta et al.~\cite{FernandezAntaMoMu13,fernandez-anta:unbounded} with their algorithm \textsc{One-fail Adaptive}, which succeeds with probability $1-O(1/n)$, again for the case with  no-CD and no knowledge of system size. The authors tightly analyze both  \textsc{One-fail Adaptive} and STB, yielding respective makespan guarantees of $ 2(\delta+1)n+O(\log^2 n)$ and $4(1 + 1/\delta)n$, where $\delta$ is a small positive constant. 

Anta and Mosteiro~\cite{Anta2010} look at a formulation of the static case---referred to as static $k$-selection (see Section~\ref{sec:deterministic})---where an unknown $k$ out of $n$ \things must each succeed. For an error parameter $\epsilon \leq 1/(n+1)$, the authors give a randomized algorithm that with probability $1-O(\epsilon)$ has makespan $O(k + \log^2(1/\epsilon))$. 

For the special case where $n=2$, Wang~\cite{dingyu:optimal} provides exact, optimal results. Specifically, the  average latency until both \things succeed, the  latency until the first \thing succeeds, and the latency until the second (last) device succeeds are shown to have expected values, respectively, of $\sqrt{3/2} + 3/2$, $2$ and approximately $3.33641$.

These results settle the question of optimal makespan for the static case using randomized approaches. 


\subsection{Deterministic and Hybrid Approaches}\label{sec:deterministic-static}

In contrast to randomized backoff, there are a handful of results that propose deterministic approaches for what is, essentially, the case of static-arrivals. Specifically, packets may arrive over time. However, once a collision occurs, those \things whose packets collided must execute a ``conflict resolution'' algorithm, while all other \things learn about the collision via channel feedback and refrain from transmitting until the conflict is resolved. 

For these results, there are $k\geq 2$ \things, each with a unique ID in the range $\{1, ..., n\}$.  Koml\'{o}s and Greenberg~\cite{komlos:asymptotically}  show the existence of a non-adaptive, deterministic algorithm that, with knowledge of $k$ and $n$, has a makespan of $O(k + k\log(n/k))$; their approach requires that each \thing $u$ only adapts to channel feedback during slots in which $u$ transmits. Related to this, Kowalski~\cite{DBLP:conf/podc/Kowalski05} proves the existence of a non-adaptive deterministic algorithm with $O(k\log(n/k))$ makespan; this is comparable to the lower bound by Greenberg and Winograd~\cite{greenberg1985lower} of $\Omega(k\log(n)/\log k)$, which applies to adaptive algorithms with collision detection.

Kowalski~\cite{DBLP:conf/podc/Kowalski05}  also gives an {\it explicit} construction for a non-adaptive algorithm with $O(k\,\texttt{polylog}(n))$ makespan. These upper-bound results stand in contrast to prior work by Capetanakis~\cite{capetanakis:generalized} and Tsybakov and Mikhailov~\cite{tsybakov1978free-russian,tsybakov1978free} who give deterministic algorithms with the same asymptotic makespan without needing to know $k$; however, their algorithms require that stations receive feedback even in slots where they themselves are not transmitting. Related to these results, 
Greenberg, Flajolet, and Ladner~\cite{DBLP:journals/jacm/GreenbergFL87,DBLP:conf/focs/GreenbergL83} give a randomized algorithm to estimate the number of  stations, which relies on collision detection; this is combined with prior deterministic  approaches~\cite{capetanakis:generalized,hayes:adaptive,tsybakov1978free-russian,tsybakov1978free,fayolle1982capacity,hofri1984stack}, yielding several hybrid algorithms, and the best  expected makespan is at most $2.1338n + o(n)$.

{\revfontg \subsection{The Wakeup Problem.}\label{sec:static-wakeup}

We introduce the \defn{wakeup problem}, which is described as follows. There are $n$ \things, of which some are active (or ``awake''), and the remainder are inactive (or``sleeping'') with only the ability to listen. Each awake \thing has a packet to send, and upon the first success, all sleeping \things ``wake up'' and the problem is solved. Typically, the metric of interest is the number of rounds until the first success. Thus, contention resolution differs from wakeup: the former requires all \things to succeed, while the latter hinges on just the first success.

Despite this difference, historically the wakeup problem has considerable overlap in the literature with contention resolution, and it is sometimes referred to by the same name (e.g., \cite{fineman:contention,fineman:contention2,gilbert2021contention,DEMARCO20171,de2013contention}). To continue the naming complications, in the {\it static} case where the set of active participants are fixed from the start, the wakeup problem is sometimes referred to as \defn{leader election (LE)}. 

Static wakeup and LE may be viewed as a natural starting point for solving contention resolution, as repeatedly applying them over all $n$ packets offers a solution, albeit an inefficient one. A more meaningful connection can be drawn between static wakeup and contention resolution: wake-up protocols often produce an estimate of  $n$, which may be used to bootstrap a contention-resolution protocol.  More specifically,
many algorithms are ``uniform'' or `fair'': in any fixed round, each \thing sends packets with the same probability (independently of the prior communication history). For such algorithms, the reciprocal of the sending probability when the first success occurs may yield an accurate estimate of $n$; obtaining system-size knowledge is a well-known component of solutions to contention resolution (e.g., \cite{jurdzinski:energy,chang:exponential-jacm,gilbert2021contention}). For instance, such an estimate can be used to set the first window size in algorithms such as \fb and \stb/\tstb  (recall Section \ref{sec:backoff-backon}), bypassing much of the backing off that would otherwise be required, and speeding up the runtime by a constant factor.

For static wakeup, the makespan refers to the number of slots until the first success (recall Section \ref{sec:metrics}). An early result on the wakeup problem is by Willard~\cite{willard:loglog}, who considers the case with CD. Here, Willard establishes a lower bound on  uniform algorithms---for any fixed round, each \thing sends packets with the same probability (independently of
the prior communication history)---{\revfontg showing that the expected makespan is at least $\lfloor\lg\lg n \rfloor - O(1)$, and an  upper bound of $\lg\lg n + O(1)$ for  expected makespan is also given.}

In the case of no-CD and where $n$ is known, 
Kushilevitz and Mansour~\cite{kushilevitz1998omega,kushilevitz1993omega} prove a lower bound of $\Omega(\log n)$ expected makespan; this applies even to unfair algorithms. An asymptotically-matching upper bound on expected makespan, and a bound of $O(\log^2 n)$ \whp is achieved by Bar-Yehuda et al. \cite{bar1992time}. In subsequent work that unifies many prior lower-bound arguments, Newport \cite{newport2014radio,newport:radio-journal} also shows $\Omega(\log n)$ expected makespan, and $\Omega(\log^2 n)$ makespan \whp for any randomized algorithms; these bounds hold even with multiple channels and/or CD.  For algorithms that succeed with probability at least $1-\epsilon$, this lower bound was strengthened to $\Omega(\log n\,\log(1/\epsilon))$ by Farach-Colton et al.~\cite{farach-colton:lower-bounds}. We note a couple things. First, these lower bounds are dramatically different from the lower bound by Willard~\cite{willard:loglog}, where CD is available. Second,  Jurdzi{\'n}ski and  Stachowiak~\cite{jurdzinski2002probabilistic}  demonstrate an algorithm with a matching upper bound of $O(\log n\,\log(1/\epsilon))$ makespan, so long as devices have unique labels, or $n$ is known.

In a setting with $\mathcal{C}> 1$ channels and CD, Fineman et al.~\cite{fineman:contention2}  give an algorithm  guaranteeing \whp $O( (\log n/\log \mathcal{C})$ $+ (\log\log n)(\log\log\log n))$ makespan  for the case where any number of \things may start in the active state,  along with an algorithm guaranteeing \whp $O((\log n/\log \mathcal{C}) + (\log\log n))$ makespan in the restricted case where at most two \things become active; this second case matches the lower bound provided by Newport~\cite{newport2014radio,newport:radio-journal}.  The authors also note that, while their result is described for the static case, it holds for the dynamic case. 

The \defn{signal-to-interference-plus-noise ratio (SINR)} model has been considered in the context of several fundamental wireless-communication problems (see \cite{moscibroda2006complexity,jurdzinski2012distributed,jurdzinski2013distributed,moses2020deterministic,jurdzinski2015cost,JurdzinskiR17}). Informally, under the SINR model,  the ratio of (i) the power of an incoming transmission to (ii) the total interference from other devices plus the ``background'' (or floor) noise, must exceed a threshold value in order for the transmission to succeed.  In this setting,  Fineman et al.~\cite{fineman:contention} show how to solve wakeup with $O(\log n + \log R)$ makespan \whp, where $R$ is ratio between the longest and shortest link. The authors point out that, in many practical settings, $R$ is at most a polynomial in $n$, which yields an $O(\log n)$ result. Also given is a general lower bound (i.e., not restricted to the class of fair algorithms) of $\Omega(\log n)$.

 
Finally, a more recent result by Gilbert et al. \cite{gilbert2021contention} examines the implications of system-size ``predictions''; that is, the algorithm is given access to the probability distribution defined over the possible values of $n$. 
The authors derive lower and upper bounds for  both deterministic and randomized solutions, for the CD and no-CD cases. Interestingly, in addition to new lower bounds, this work recovers some familiar results, such as the lower bound by Willard~\cite{willard:loglog}.
}

\subsubsection*{\bf Summary.} For static contention resolution, many randomized algorithms have been proposed, with STB providing one of the strongest results: asymptotically-optimal $O(n)$ makespan with no-CD,  no knowledge of $n$,  and only $O(\log n)$ channel-access attempts.  In contrast, purely deterministic approaches cannot achieve linear makespan, and these results rely on additional assumptions regarding channel feedback, knowledge of $n$, or a unique labeling of \things.  

In the context of network traffic, we highlight the underlying assumption that each transmission of a packet is accommodated by a single slot. Unsurprisingly, \beb also has sub-optimal makespan for the more general case with heterogeneous packet sizes, where a packet of size $s$ must have sole access to the channel for $s$ consecutive slots in order to succeed~\cite{bender:heterogeneous}. Along similar lines, in the wireless domain, it has been argued that packet size should be incorporated into the definition of makespan, since in WiFi networks collisions may consume time proportional to packet size~\cite{anderton:is,anderton:windowed}. 

{\revfontg For static wakeup, under each of CD or no-CD, asymptotically matching upper and lower bounds are known. It is interesting to compare these results to those of static contention resolution. In particular, while the makespan for contention resolution is higher, the optimal average per-packet time to success is $\Theta(1)$, which is lower than even the CD case for wakeup.}

\section{Dynamic Arrivals}\label{sec:adversarial-dynamic}

What can be said about the case where packets do not arrive and succeed in disjoint batches, but rather are injected into the system over time? A large body of literature addresses contention resolution where arrival times are governed by a stochastic process, typically Bernoulli or Poisson. Informally, with a stochastic setting, the aim is to show that packets do not grow without bound at each sender, but are successfully delivered at a pace sufficient to keep up with the rate that new packets are generated. For the interested reader, we highlight the survey by Chlebus~\cite{chlebus2001randomized}, which addresses in depth the literature in this area.

However,  modeling of arrival times via a stochastic process may not always be justified.  In particular, bursty traffic is common in practice~\cite{williamson:internet,wilson1991high,paxson:wide-area,Teymori2005,yu:study}, and its impact has been examined~\cite{Ghani:2010,sarkar:effect,canberk:self,bhandari:performance}. Indeed, the static case discussed previously in Section~\ref{sec:backoff-algorithms} illustrates the impact on makespan by disjoint bursts of packets.  This has led the research community to consider more general arrival models.

To this end, adversarial queuing theory (AQT)~\cite{borodin2001adversarial,cruz-one,cruz-two,andrews2001universal} has been used to model packet arrivals times, and many contention-resolution results fit within the AQT framework; that is, they address adversarial models that are parameterized by rates of packet injection and burstiness. Another large body of the contention-resolution literature eschews the parameterization of AQT and instead considers a completely arbitrary schedule for when packets arrive. Here,  we treat separately the literature on deterministic and randomized approaches, in  Sections~\ref{sec:deterministic} and~\ref{sec:random-stream}, respectively.

\subsection{AQT and Contention Resolution}\label{sec:aqt-results}

In this setting,  each \thing is sometimes associated with a queue into which packets may be injected by an adversary at arbitrary times. Other results are presented in a queue-free model, where each packet has a corresponding station that ``dies'' (i.e., becomes irrelevant) after the packet is successfully transmitted (see discussion by Goldberg~\cite{Goldberg-notes-2000,anantharamu2010deterministic}). In either case, packet arrival times are parameterized by an injection rate  {\boldmath{$\rho$}}, where $0<\rho \leq 1$, and a non-negative integer, {\boldmath{$b$}}, that denotes burstiness. 

Multiple adversarial models appear in the AQT literature; see Borodin~et al. \cite{borodin2001adversarial} for an in-depth discussion. In the context of contention resolution, there are two prominent models. The \defn{leaky-bucket adversary} is defined such that, for any $t$ consecutive slots, at most $\rho t + b$ packets may be placed in the queue of  any \thing~\cite{cruz-one}. For a contiguous segment of rounds, and any positive integer $w$, a \defn{window adversary} of type $(\rho,w)$ can inject $\rho w$ packets in each contiguous segment of $w$ rounds into any set of stations; $w$ is referred to as a window.

Performance is typically measured by the growth of queues at each \thing. As defined by Andrews et al.~\cite{andrews2001universal},  an algorithm is \defn{stable} on a network  despite an adversary if, starting from a state where there are no packets, the total number of packets is always bounded by a constant. An algorithm is \defn{universally stable} if it is stable  against every adversary with $\rho < 1$, for any network~\cite{andrews2001universal}. Finally, in the contention-resolution literature, the notion of \defn{strongly stable} arises in the context of a window adversary: an algorithm is \defn{strongly stable} if, at any time,  the total number of queued packets is $O(\rho w)$~\cite{chlebus2006adversarial,chlebus:adversarial}. 

Another important metric is (packet) \defn{latency}, which is often defined as the maximum time that any packet stays in the queue prior to being successfully transmitted. We note that, while this is described in the context of queues, this aligns with the metric as described in Section~\ref{sec:metrics}.

\begin{figure}[t]
\centering
\includegraphics[scale=0.4]{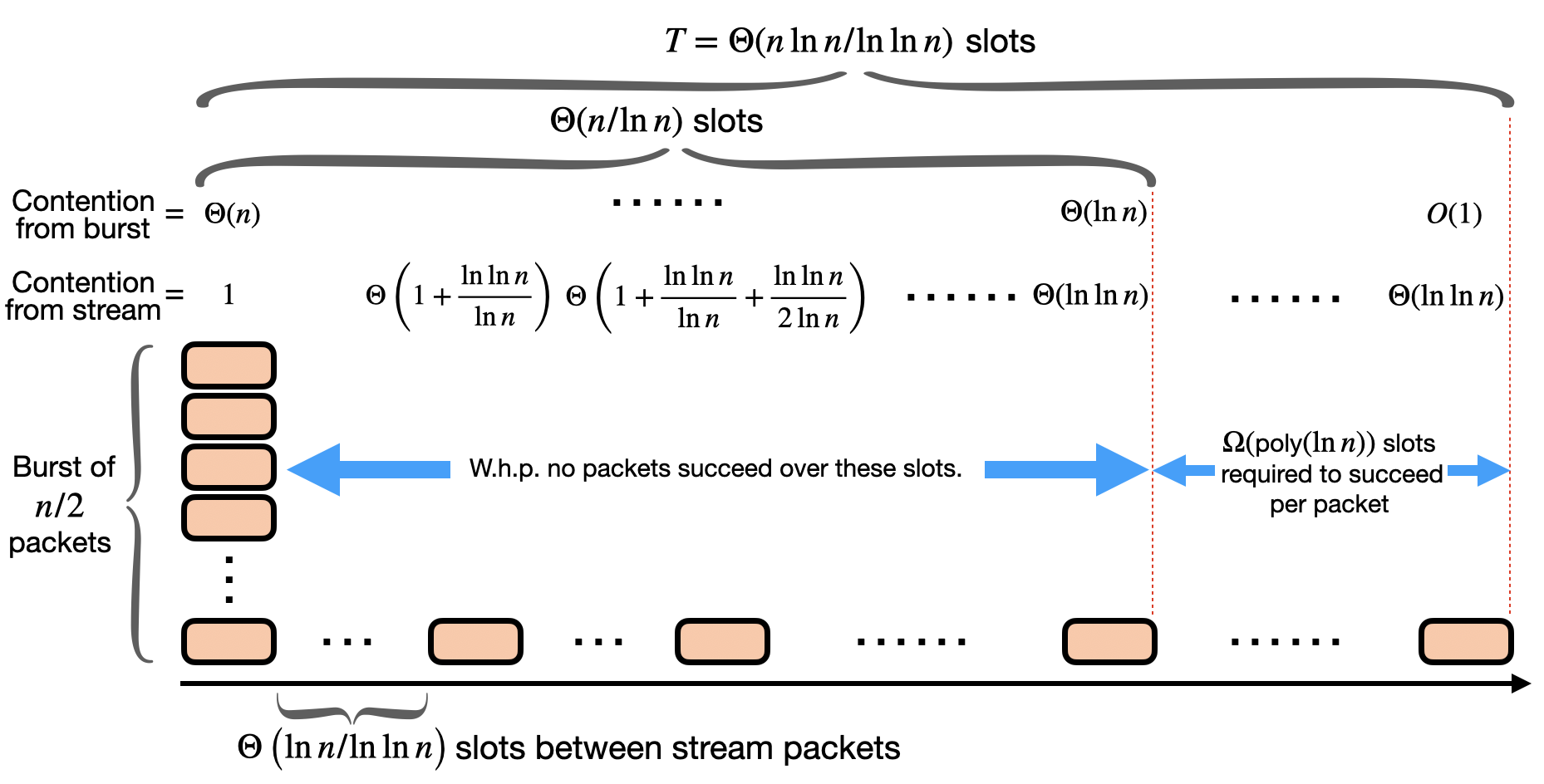}
\vspace{-5pt}\caption{Depiction of how exponential backoff is unstable for a packet arrival rate of $\Omega(\ln\ln n/\ln n)$.}\label{fig:aqt-intuition} \vspace{-10pt}
\end{figure}

\subsubsection*{\bf Developing Intuition.}\label{s:intuition-aqt} Given the prominence of backoff algorithms, one may wonder about their performance in the AQT setting. Given the suboptimal makespan of popular algorithms like BEB in the static case, we should be skeptical of their performance when packets arrive dynamically at a high rate. To further this intuition, we sketch an argument that shows the instability of an algorithm \beb', which is similar to \beb. Specifically, for each slot $i\geq 1$ starting with the slot that packet $u$ injected into the system, $u$ sends with probability $1/i$. Note that this is  similar to the Bernoulli-backoff algorithm (recall Section~\ref{sec:backoff}), where the sending probability per slot proceeds as $1/2, 1/2, 1/4, 1/4, 1/4, 1/4, 1/8, 1/8, ...$ etc. By comparison, \beb' proceeds as $1, 1/2, 1/3, 1/4, 1/5, ...$, and so at any given slot since a packet is injected, the sending probabilities under the Bernoulli-backoff algorithm and \beb' differ by only a  constant.\footnote{We highlight that this is very similar to the algorithm \textsc{Decrease Slowly} given by 
Jurdzi\'nski and Stachowiak \cite{jurdzinski2005probabilistic,jurdzinski2002probabilistic}; see the proof of Theorem 8 in \cite{jurdzinski2005probabilistic} for a formal analysis.}

We can now give some intuition for why \beb' is unstable under an arrival rate of $\Omega( \ln\ln n/\ln n)$;  Figure~\ref{fig:aqt-intuition} complements our discussion. Consider an interval of $T=\Theta(n \ln n/\ln\ln n)$ slots. In the first slot,  a single burst of $n/2$ packets arrives. This drives up the contention to $\Omega(n)$ in this slot. None of these packets is likely to succeed over the next $n/\ln n$ slots; to see this, note that in the $i$-th slot measured from this first slot is sending with probability $1/i$, yielding a contention of $\Omega(\ln n)$ in each of these $n/\ln n$ slots, and so the probability of success is $O(1/\poly{n})$  by Equation~\ref{eq:success-prob}.

In addition to the burst, there is a steady stream of $n/2$ packets: a single packet is injected every $\Theta(\ln n/\ln\ln n)$ slots measured after the first slot. Given our discussion in Section~\ref{sec:backoff}, by itself, this stream poses no problem for \beb'. Specifically, by the lower bound in Equation~\ref{eq:success-prob}, we expect each packet to succeed within a small constant number of slots prior to the next packet arriving and, therefore, contention remains low.

However, given the burst of packets, the packets in the stream face $n/\ln n$ slots where \whp no successes can occur. Consequently, these packets accumulate in the system and, over these $n/\ln n$ slots, their contribution to contention becomes $\Omega(\ln\ln n/\ln n)\sum_{i=1}^{n/\ln n} 1/i = \Omega(\ln\ln n)$ by the sum of a harmonic series. By Equation~\ref{eq:success-prob}, the probability of success is $O(1/\poly{\ln n})$, and thus the expected number of slots for a packet to succeed is $\Omega(\poly{\ln n})$. 

To summarize the above situation, we expect that for all $n$ packets to succeed  $\Omega(n\cdot \poly{\ln n})$ slots are required, which exceeds $T$. Consequently, for every such interval $T$, more and more packets accumulate in system over time.  Note that this instability occurs for a rate that is $n/T = \Omega( \ln\ln n/\ln n)$. The above argument originates with Bender et al.~\cite{bender:adversarial}, who use  it to establish bounds on the arrival rate for which several backoff algorithms are unstable. 

\medskip

\noindent{\bf Overview of Literature.} As mentioned above, the result by Bender et al.~\cite{bender:adversarial} explores stability of backoff algorithms under windowed adversaries in the queue-free model. \beb is shown to be stable for $\rho = d/\log n$ for some constant $d>0$, and unstable for $\rho = \Omega(\log\log n/\log n)$ . They also show that \llb is stable for $\rho=d'/( (\log n)\log\log n)$ for some constant $d'>0$, and unstable for $\rho = \Omega(1/\log n)$. Overall, traditional backoff algorithms appear to perform sub-optimally, although this is not surprising given the corresponding makespan results for the static case.


The remaining approaches in the literature that we survey are deterministic. Chlebus et al.~\cite{chlebus2009maximum,chlebus2007stability} consider adversaries where $\rho=1$; that is, one packet is injected per slot. The main positive result is a deterministic, adaptive protocol that guarantees against leaky-bucket adversaries that $O(n^2 + b)$ packets are queued in the system at any given time, where $n$ is the number of \things. Here, $n$ is known to all \things and each ID is unique in the range $[1, ..., n]$; this allows \things to schedule via passing a token, such that no collisions occur (notably, the positive results in this paper do not require collision detection). This result is shown to be asymptotically optimal in the sense that  a leaky-bucket adversary with burstiness $2$ can force $\Omega(n^2)$ total packets to be queued.

Several impossibility results are also given---holding regardless of whether or not  collision detection is available---that illustrate how even minor burstiness can prevent stability. Specifically, no full-sensing protocol can be stable  (i) for at least two \things against a leaky-bucket adversary with burstiness $2$, (ii) for at least three \things against a window adversary of burstiness $2$; and (iii) no acknowledgment-based algorithm is stable in a system of at least two \things against the window adversary of burstiness $2$.

Additional impossibility results are made in the context of three properties. First, an algorithm is \defn{fair} if each packet is transmitted successfully within $O(b/\rho)$ slots from the time it is injected (this is a different notion of fairness than described in Section~\ref{sec:static-wakeup}).
 The authors show that  no algorithm can be fair {\it and} stable for any number of at least two \things against a leaky-bucket adversary of burstiness $2$. 
Second, an algorithm is \defn{withholding} the channel if, upon successfully transmitting a packet, the station keeps sending until all of its pending packets are exhausted. No algorithm that withholds the channel can be stable for three \things against a leaky-bucket adversary of burstiness $2$. Third, an algorithm is \defn{retaining} if after a \thing, say $u$,  successfully transmits a packet, and if any packets arriving subsequent to this event are injected only at other \things, $u$ eventually empties its queue. No retaining protocol is stable with at least four \things against a window adversary of burstiness $2$. Chlebus et al. show that this implies no fair protocol can also be stable with at least four \things against the window adversary of burstiness $2$. 


In contrast to the above,  Chlebus et al.~\cite{chlebus2006adversarial,chlebus:adversarial} study deterministic algorithms under the same model as Chlebus et al.~\cite{chlebus2009maximum,chlebus2007stability}, but where  $\rho<1$. They show that strong stability is impossible for a rate $\omega(1/\log n)$, with or without CD. With CD, the authors give a full-sensing algorithm that is fair and stable for $\rho = O(1/\log n)$, and they show that fairness is not possible for rates that are $\omega(1/\log n)$. When collision detection is not available,  the authors give an existence proof for a stable and fair algorithm under a much smaller rate of $O(1/n\log^2 n)$, and they give an explicit algorithm that achieves these properties for a rate of $O(1/n^2\log n)$. Finally, it is shown that no acknowledgement-based algorithm can be stable for a rate exceeding $3/(1+\lg n)$. 


Anantharamu et al.~\cite{anantharamu2010deterministic} and Aldawsari et al.~\cite{aldawsari:adversarial} design and analyze several deterministic algorithms that employ either a round-robin or token-passing approach for obtaining access to the channel. The authors derive bounds on latency under the leaky-bucket adversary model as a function of $\rho$,  for $\rho < 1/(2\log n)$, and general $b$. Note that this restriction on $\rho$ should not surprise us, given the negative results on strong stability by Chlebus et al.~\cite{chlebus2006adversarial,chlebus:adversarial} and prior results on the challenges posed by burstiness~\cite{chlebus2009maximum,chlebus2007stability}. These algorithms are compared, both analytically and via simulations, against the randomized algorithms, BEB and Quadratic Backoff (QB), where the latter has each successive window size increase as the square of the previous window. The simulation results illustrate the impact of different injection rates, burstiness values, and the number of \things. These findings generally align with the theoretical predictions and, interestingly, the deterministic algorithms compare favorably to BEB and QB, suggesting that randomized approaches may not necessarily be superior. More-recent work by Hradovich et al.~\cite{HradovichKK21} provides a new average-case analysis of the maximum queue size and latency for several of the same algorithms.



While many deterministic approaches require nodes to have unique IDs,  Anantharamu and Chlebus~\cite{anantharamu2015broadcasting} consider deterministic algorithms  where this is not true; furthermore, $n$ is unknown to the algorithm. In trade, they focus on positive results against leaky bucket adversaries that have the additional constraint that they inject at most one packet per round. An algorithm is \defn{activation-based} if stations that do not have any packets to send ignore any channel feedback. In terms of positive results, the authors provide a non-adaptive, activation-based algorithm that has bounded latency and is stable for  an injection rate $\rho<1/3$ with CD, a non-adaptive full-sensing algorithm that has bounded latency and is stable for $\rho \leq 3/8$ with CD, and an adaptive activation-based algorithm that has bounded latency and is stable for injection rates less than $\rho \leq 1/2$ with no-CD.

In terms of negative results, Anantharamu and Chlebus show that no deterministic algorithm can be fair against an adversary that is able to activate at most $2$ stations in any round---called a $2$-activating adversary---with $\lfloor \rho + b \rfloor = 1+b\geq 2$.   Additionally,  the authors show that no deterministic acknowledgment-based algorithm is fair
against a $1$-activating adversary for $\rho<1$ and $b\geq 3-2\rho$. This stands in contrast to the positive results in Chlebus et al.~\cite{chlebus2006adversarial,chlebus:adversarial} for $\rho<1$. Finally, Anantharamu and Chlebus prove that no deterministic algorithm with CD can provide bounded latency against a $1$-activating adversary with injection rate at least $3/4$ and $b \geq 1$.  It is interesting to compare this result against the algorithm by Chlebus et al.~\cite{chlebus2009maximum,chlebus2007stability} discussed above, which achieves stability, having $O(n^2 + b)$ total packets queued in the system at any given time (but where there are unique IDs and the number of stations is known).


Anantharamu et al.~\cite{anantharamu2009adversarial,anantharamu2017adversarial} consider deterministic full-sensing and acknowledgment-based algorithms for the case where each \thing is subject to an individual injection rate under a window adversary, where the window size $w$ is unknown. There are $n$ \things, each with a unique ID in ${1,..., n}$, and $n$ is known. The authors prove a lower bound on latency of $\Omega(w \max\{1, \log_w n\})$ for $w\leq n$, and $\Omega(w+n)$ for $w>n$. Upper bounds are established by a full-sensing algorithm with CD, and by an adaptive algorithm without CD, each achieving a latency of $O(\min(n + w, w \log n))$; Anantharamu et al. also give a full-sensing algorithm without CD that has $O(n + w)$ queue size  and $O(nw)$ latency.



Bienkowski et al.~\cite{bienkowski:distributed,BienkowskiJKK12} consider a model where $n$ is known, IDs are unique, each \thing has a queue, and messages may include control bits. The adversary considered here differs from the leaky-bucket and window adversaries discussed earlier: at any time, an arbitrary number of packets can be injected into the queue of any \thing. The metrics of interest are the total packets in all queues, called \defn{total load}, and the maximum queue size.  The authors prove that their deterministic algorithm achieves a total load that is an additive $O(n^2)$ amount larger than the best offline algorithm, and the maximum queue size is at most an $n$-factor larger. 


The remaining results assume assume a model where $n$ is known, and \things have unique IDs and access to a global clock. Hradovich et al.~\cite{hradovich2020contention} consider a problem variant where, in any fixed round, at most $k$ stations may access  the multiple access channel; this is called a {\boldmath{$k$}}\defn{-restrained channel}. The authors explore deterministic algorithms for leaky bucket adversaries with $\rho\leq 1$. The first is a stable adaptive algorithm whose  queue size is $O(n^2 + b)$, which is reminiscent of the bound in \cite{chlebus2009maximum}. The second is a full-sensing algorithm that is stable for $\rho\leq 1-1/n$. Finally,  the authors give an acknowledgment-based algorithm with throughput $\Theta(k/n(\log^2n))$, which is complemented by a lower bound that no acknowledgement-based algorithm can achieve a throughput better than $\min\{k/n, 1/(3\log n)\}$. Simulations indicate that the proposed algorithms perform favorably, with respect to queue size and the number of channel accesses, relative to well-known backoff algorithms.


In work by Garncarek et al.~\cite{garncarek2018local}, \things receive no channel feedback, and they may inform their decisions using knowledge of their own local queue, or information gained about other queues. In particular, an adaptive algorithm may store bits in memory that reflect queue histories, while non-adaptive (memoryless) algorithms may not. Both types of algorithms can check whether or not their queue is empty. The authors design a deterministic, adaptive algorithm that guarantees that all stations have bounded queue size $L_{\mbox{\tiny max}}$ for any $\rho$ and $b$; that is, the algorithm is universally stable. The amount of state that each \thing  maintains is at most $n \log(L_{\mbox{\tiny max}})$ bits. The authors also prove that there is no non-adaptive algorithm that is stable for an injection rate of $O(1/\log n)$, and they give a non-adaptive algorithm that is stable for an injection rate of $d/\log^2n$, for some constant $d>0$.


More recently,  Garncarek et al.~\cite{garncarek2019stable} operate in the same model, with a focus on algorithms that keep very little or no information about the previous state of the system. Here, the authors show a deterministic algorithm that is stable for $\rho = \Omega(1/\log n)$ given that only $O(1)$ bits are used as memory to store some history of events, and each \thing can check its queue size (not just whether it is empty). Stability can also be achieved when the rate is increased to a constant, and with zero bits of memory; however, $b$ must be upper bounded by some value known {\it a priori}.


\medskip

\subsubsection*{\bf Summary.} There are several significant messages conveyed by the literature surveyed above. Randomized backoff algorithms continue to disappoint in that they are stable only for sub-constant injection rates. Interestingly, in terms of latency, deterministic algorithms can achieve performance that rivals and even surpasses traditional randomized backoff algorithms.  However, most of these results make assumptions about unique IDs and/or {\it a priori} knowledge of $n$ (an exception is the algorithm by Anantharamu and Clebus~\cite{anantharamu2015broadcasting},  which assumes a $1$-activating adversary). This leads to a more general message: while stability and fairness seem like natural qualities that we desire in algorithm design, it is often not possible to achieve both simultaneously when there is even a little (i.e., constant) burstiness. Therefore, all of this paints a fairly dismal picture of what is possible under window and leaky bucket adversaries, given the many impossibility results that apply.

What avenues are there for positive results? As we will discuss later in Section~\ref{sec:random-stream}, new randomized approaches offer better performance.  However, fundamentally, perhaps the AQT model is overly pessimistic; this has been investigated in more general settings (e.g., \cite{BKS14}.) Here, focusing on whether queue sizes can become unbounded under a very specific injection pattern, and under an infinite stream of packets, may lead to results that are artificially bleak. In the next section, we survey work that is concerned with a finite stream of packets that arrive under an arbitrary schedule.


\subsection{Deterministic Approaches}\label{sec:deterministic}


\begin{itemize}[leftmargin=5.85cm]
\item[]{\it ``Anyone who attempts to generate random numbers by deterministic means is, of course, living in a state of sin.''} \hspace{3pt} -- John von Neumann
\end{itemize}
\vspace{10pt}

In this section, we depart from parameterized adversaries and notions of stability, while still focusing on deterministic contention-resolution algorithms. Much of the contention-resolution work in this setting is referred to as  {\boldmath{$k$}}\defn{-selection}, which is described as follows. There are $n$ \things of which $k$, for $1 \leq k\leq n$, activate over time, each with a {\it single} packet to transmit.  The activation times are selected by an adversary. As before, most of the literature operates in a model where each \thing has a unique ID.  Furthermore, $n$ is typically assumed to be known; throughout this section, we assume this to be true unless explicitly stated otherwise. The goal is to minimize the latency (recall from Section~\ref{sec:metrics}) required for all $k$ \things to successfully send their respective packets.

\begin{figure}[t]
\centering
\includegraphics[scale=0.45]{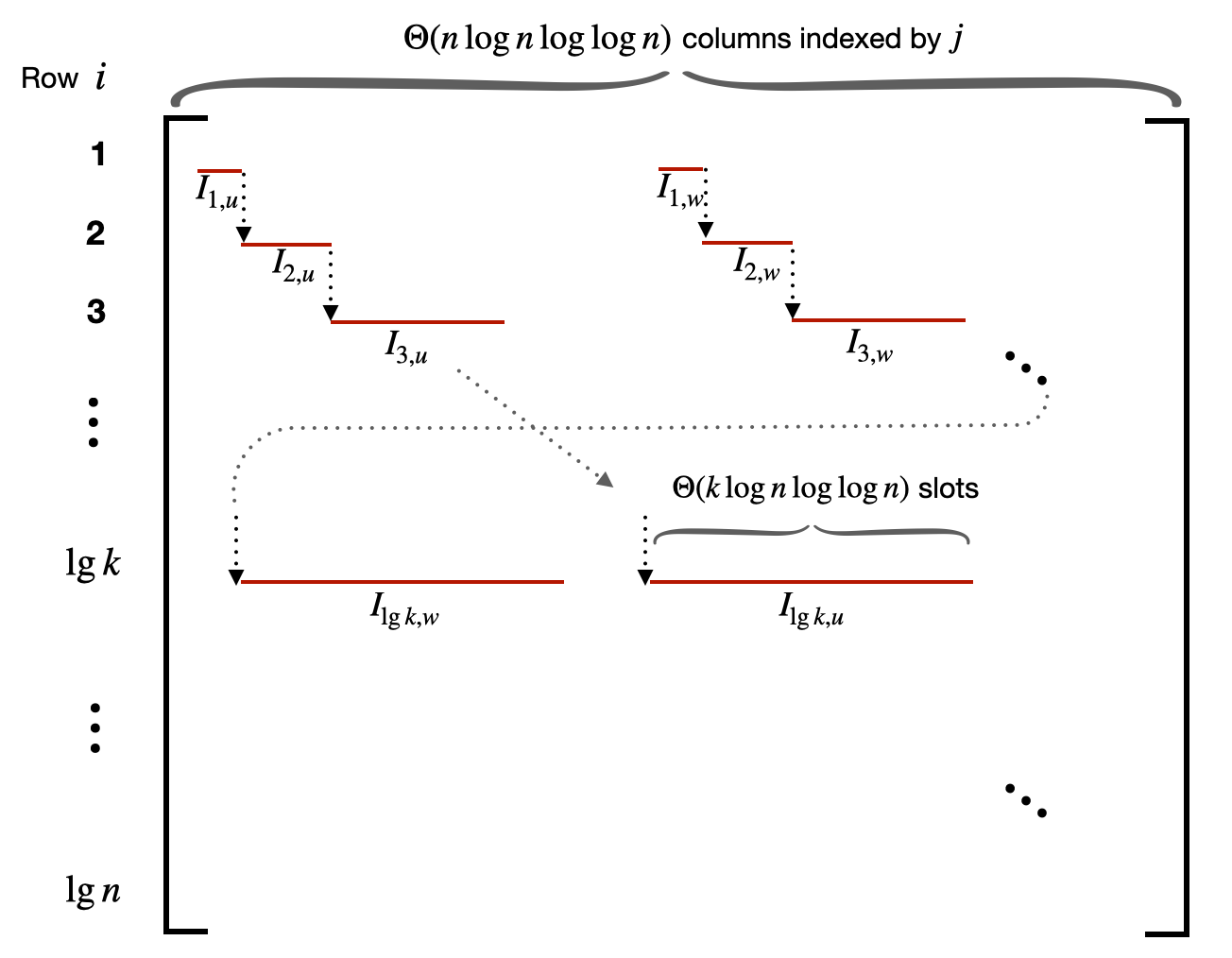}
\vspace{-5pt}\caption{Depiction of a transmission matrix presented in the work by De Marco and Kowalski~\cite{doi:10.1137/140982763}. For two \things, $u$ and $w$, we illustrate the intervals over which they respectively may transmit on the shared channel. }\label{fig:det-intuition}
\end{figure}

\subsubsection*{\bf Developing Intuition.} Many results put a spin on John von Neumann's quote by generating deterministic results from randomized approaches. Specifically, a common technique in this area is to design a randomized algorithm, and then use the probabilistic method~\cite{alon2016probabilistic} to argue that a deterministic analog  exists.  Here, we focus on the result by De Marco and Kowalski~\cite{doi:10.1137/140982763}; however,  related approaches are used in several other papers to give existence proofs for efficient transmission schedules~\cite{DBLP:conf/icdcs/MarcoKS19,de2013contention,DEMARCO20171,Chlebus:2016:SWM:2882263.2882514,ChlebusK04,ChlebusGKR05}.
De Marco and Kowalski~\cite{doi:10.1137/140982763} consider the dynamic $k$-selection problem  where there is a global clock, $k$ is unknown, and no-CD. Their main result is a non-adaptive, deterministic algorithm for contention resolution with latency $O(k\log(n)\log\log(n))$. 

A key component of the authors'~\cite{doi:10.1137/140982763} approach is a \defn{transmission matrix}, $\mathcal{M}$, consisting of $\lg n$ rows and $\ell=\Theta(n \log n \log\log n)$ columns. The way in which \things use  $\mathcal{M}$ to transmit on the channel is as follows. Let $\mathcal{M}_{i,j}$ denote the entry at row $i$ and column $j$; each such entry is a list of IDs. Upon activating, in each slot \thing $u$ scans left-to-right in row $1$ of $\mathcal{M}$. In this row, there is a designated interval of slots $\mathcal{I}_{1, u}$ for $u$; specifically, $u$ rounds up the current (global) slot index to the nearest multiple of $2\lg\lg n$ and this is the start of $\mathcal{I}_{1, u}$, which has length $\Theta(\log n \log\log n)$.
If $u$ belongs to the list at $\mathcal{M}_{1,j}$ for $j\in \mathcal{I}_{1, u}$, then $u$ sends its packet in slot $j$. If at any point $u$ succeeds in $\mathcal{I}_{1, u}$, then $u$ terminates; otherwise, $u$ immediately proceeds to row $2$ and starts executing as prescribed by $\mathcal{I}_{2, u}$. This process through $\mathcal{M}$ continues until all rows are exhausted. The length of interval, $\mathcal{I}_{i, u}$, in each row $i$ is $\Theta(2^i \log n\log\log n)$.  Figure~\ref{fig:det-intuition} illustrates a transmission matrix.
 
How is membership in $\mathcal{M}_{i,j}$ determined? This is decided {\it randomly}: \thing $u$ belongs to  $\mathcal{M}_{i,j}$ with probability approximately $\frac{1}{2^i 2^{j\hspace{-2pt}\mod \lg\lg n}}$, for $1 \leq i\leq \lg n$ and $j\geq 1$. 
Note that the first term of the sending probability (the $1/2^i$ term) and also the interval length (specifically, the $2^i$ term) are reminiscent of the $i^{\mbox{\tiny th}}$ window for the Bernoulli version of \beb (recall Sections~\ref{sec:backoff} and \ref{sec:aqt-results}). Consequently, we should not be surprised at a $\Omega(k \log n)$ latency.

How does contention behave under $\mathcal{M}$? For illustrative purposes, consider $\Theta(k)$ \things executing the schedule in rows $1$ and $\lg k$. In row $1$, an active \thing $u$ may contribute significantly to contention, but only for its small interval $\mathcal{I}_{1, u}$. For example, if $\Theta(k)$ \things all activate simultaneously (i.e., a batch), then their contribution to contention is $\Theta(k)$ and no \thing succeeds; however, this occurs only over $\Theta(\log n \log\log n)$ slots in row $1$. Conversely, the $\Theta(k)$ \things may be spread out over row $1$, each contributing only $O(1)$ contention in their own respective, disjoint interval of $\Theta(\log n \log\log n)$ slots; we will return to these two extremes momentarily.

In row $\lg k$, the sending probability ranges between $\Theta(1/k)$ and $\Theta(1/(k\log n))$. Thus, a batch of $\Theta(k)$ \things make a more muted contribution to contention, ranging from $\Theta(1)$ to $\Theta(1/\log n)$ over a much larger interval $\mathcal{I}_{\lg(k)}$ consisting of $\Theta(k\log n\log\log n)$ slots. This range of contention is ``good'' in the sense that, recalling Equation~\ref{eq:success-prob}, in the absence of other major contributions to contention, we expect \things to succeed.  Is this the case?

Recall the contribution to contention from \things in row $1$ using the two qualitatively different cases discussed above:  their contribution may be high for a ``small'' number---specifically, $\Theta(\log n\log\log n)$---slots of $\mathcal{I}_{\lg(k)}$, or only $O(1)$ over perhaps all of $\mathcal{I}_{\lg(k)}$.  Therefore, accounting for these two cases, in row 1 there should be many disjoint periods of $\Theta(\log\log n)$ contiguous slots where contention is good. More generally, given the contribution to contention from other rows, the authors show that, w.h.p. in $n$, an important property holds: in every interval of length $\Theta(k\log(n)\log\log(n))$ slots,  $O(k/\log n)$ \things  proceed from row $\lg(k)$ to row $\lg(k) + 1$ of the transmission matrix. De Marco and Kowalski~\cite{doi:10.1137/140982763} show that a random transmission matrix yields the above property with strictly positive probability. Therefore, by the probabilistic method, the authors argue that there must exist a matrix with this property.

But what about these $O(k/\log n)$  \things that keep executing in row $\lg (k) + 1$? Interleaved with the execution is the deterministic result by Koml\'os and Greenberg~\cite{komlos:asymptotically} (recall Section~\ref{sec:deterministic-static}), which addresses the static case. Specifically,   given a batch of $m$ out of $n$ \things with a packet to send, where both $m$ and $n$ are known,  all $m$ \things  succeed in time $O(m + m\log(n/m))$. Here, in the dynamic $k$-selection problem, $m$ is not known, and so the result by Koml\'os and Greenberg is executed for values of $m=1, 2, ..., \lg n$, which contributes an extra logarithmic-in-$n$ factor to the run time. Thus, the $m=O(k/\log n)$ \things that remain after row $k$ succeed within an additional $O(m + m\log(n/m))\log n$, which is  $O(k \log(n\log n))$ by substituting in for $m$.

Finally, note that by the sum of a geometric series, the number of slots in $\mathcal{M}$ over which any \thing executes prior to reaching row $\lg(k) + 1$ is $O(k\log n \log\log n)$. Adding to this the $O(k \log(n\log n))$  slots for the Koml\'os and Greenberg result yields a total latency of $O(k\log n \log\log n)$.

\subsubsection{Deterministic $k$-Selection} 

We start by reiterating that Greenberg and Winograd~\cite{greenberg1985lower} give a lower bound for deterministic algorithms of $\Omega(k\log n/\log k)$ latency in the static case with CD. This bound, of course, applies to the dynamic case.  In particular, as discussed above, De Marco and Kowalski~\cite{doi:10.1137/140982763} demonstrate a non-adaptive, deterministic algorithm that achieves $O(k\log(n)\log\log(n))$ latency, which not far off of the lower bound by Greenberg and Winograd~\cite{greenberg1985lower}.



Subsequently, De Marco et al. \cite{DBLP:conf/icdcs/MarcoKS19} examine the power of acknowledgement-based versus non-acknowledgement-based deterministic non-adaptive  approaches when there is no global clock and under the no-CD setting. For unknown $k$,  they give a non-acknowledgement-based algorithm with latency $O(k^2 \log n)$, and show that this can be improved to $O(k^2 \log n/\log k)$ with an acknowledgment-based algorithm. These results are (nearly) tight; the authors also provide lower bounds on latency of $\Omega(k^2/\log k)$ for both cases.  This illustrates a startling contrast with the case for known $k$, where the authors give an acknowledgement-based algorithm with $O(k \log k \log n)$ latency, which is known to be nearly tight \cite{ClementiMS03}.


\subsubsection{Deterministic Wakeup}\label{sec:det-wakeup} 
Unlike the static version, {\it dynamic} wakeup differs significantly from LE. Here, a subset of \things become awake according to some arbitrary schedule over time, while the remainder are asleep. For consistency in discussing the dynamic setting, latency (rather than makespan) is our terminology used to report results. 

There are a number of results for the single-hop setting. To start with a challenging model, De Marco, Pellegrini, and Sburlati~ \cite{DBLP:journals/dam/MarcoPS07} examine the case of no global clock, no-CD, and where $n$ is unknown. They give a non-adaptive, deterministic algorithm with $O(n^3\log^3 n)$ latency (a bound of $O(n^4 \log ^3 n)$ is reported in an earlier version of this work). This improves over the previous best of $O(n^4\log^5 n)$ latency by G\k{a}sieniec, Pelc, and Peleg~\cite{DBLP:journals/siamdm/GasieniecPP01,DBLP:conf/podc/GasieniecPP00}.

Continuing in a model with no-CD, but with a global clock, 
De Marco and Kowalski~\cite{DEMARCO20171,de2013contention} consider non-adaptive solutions. Here, there is a parameter $k$, for $1\leq k\leq n$, which is an upper bound on the number of stations that may become active over time. Of these stations,  $s$ is the time (measured from the first slot) at which the first station becomes active. The authors  look at three settings: where $s$ is known and $k$ is unknown, when $k$ is known  and $s$ is unknown, and when both $k$ and $s$ are unknown. In the first two settings, the authors give algorithms that achieve a latency of $O(k \log(n/k) + 1)$, while their solution for the third scenario has a latency of $O(k \log(n) \log\log(n))$. A lower bound of $\Omega(k \log(n/k) + 1)$ is given, which shows that their first two upper bounds are asymptotically optimal, and the third is within a $O(\log\log n)$-factor of optimal. 

Chlebus et al. \cite{Chlebus:2016:SWM:2882263.2882514} consider a multi-channel scenario with no-CD and no global clock. 
Here, all stations may access all $b\geq 1$ channels at once, and with the same parameter $k\geq 1$ being an upper bound on the number of stations that become active over time. The authors give a non-adaptive,  deterministic algorithm for the cases where $k$ is unknown that has latency $O(k\log^{1/b}k \log n)$. When  $b=\Omega(\log\log n)$, a faster deterministic algorithm is possible, with $O((k/b) \log n \log(b\log n))$ latency, and a lower bound of $\Omega((k/b)\log(n/k))$ is shown for deterministic algorithms. The authors also consider probabilistic jamming (a topic discussed further in Section \ref{sec:jamming}), where each channel is jammed independently and uniformly with known probability $p$ per round, where $0<p<1$. Here,  Chlebus et al.~\cite{Chlebus:2016:SWM:2882263.2882514}  give two deterministic algorithms that succeed with probability $1-O(1/\texttt{poly}(n))$, the first with latency $O(k\log(n)(\log^{1/b}k)/\log(1/p))$, and the second with latency $O((k/b)\log(n)\log(b\log(n))/\log(1/p))$ when $b = \Omega(\log( b\log n))$.

Switching to the multi-hop setting, Chrobak, G\k{a}sieniec, and Kowalski~\cite{ChrobakGK07} also consider the wakeup problem with no-CD and no global clock. The authors present a deterministic algorithm with run time $O(n^{5/3}\log n)$. This was later improved by
Chlebus and Kowalski~\cite{ChlebusK04}, who give (non-constructively) a non-adaptive, deterministic algorithm with $O(n^{3/2}\log n)$ time; they also provide a construction that achieves $O(k^2 \poly{\log n})$ latency. Finally, Chlebus et al.~\cite{ChlebusGKR05} improve this further with an existence proof for a wakeup algorithm with $O(n\log^2 n)$ latency, and an explicit algorithm that runs in $O(n\Delta \poly{\log n})$ time, where $\Delta$ is the maximum in-degree.

Finally, under the SINR channel model, Jurdzi{\'n}ski and Stachowiak~\cite{jurdzinski2015cost} show that any deterministic algorithm with access to a global clock must incur $\Omega(\log^2 n)$ latency. This complements an earlier result by Jurdzi{\'n}ski et al.~\cite{jurdzinski2013distributed}, who give a deterministic algorithm with $O(\log^2 n)$ latency in the single-hop setting.



\medskip

\subsubsection*{\bf Summary.} Much of the theoretical territory has been mapped out with some nearly-matching upper and lower asymptotic bounds on latency, and the results are less bleak than those surveyed under the AQT model. Notably, with no-CD and assuming a global clock,  we see this for both $k$-selection (De Marco and Kowalski \cite{doi:10.1137/140982763}) and for wakeup (De Marco and Kowalski~\cite{DEMARCO20171,de2013contention}), where the latter result depends on some limited system information.

Despite providing a rosier picture, deterministic approaches for contention resolution fall short of being able to offer linear latency in the dynamic case; this is unsurprising given the analogous situation in the static case (Section~\ref{sec:deterministic-static}). Similarly, deterministic solutions to the wakeup problem yield high latency in cases without a global clock, or where key parameters are unknown.  A natural place to look for improvement---perhaps even asymptotic optimality---is the domain of randomized algorithms, which we survey next.

\subsection{\bf Randomized Approaches for Contention Resolution}\label{sec:random-stream}

As we saw in Section~\ref{sec:backoff-algorithms},  randomized backoff algorithms provide asymptotically-optimal makespan for the static case without requiring unique IDs, CD, or knowledge of $n$.  Thus, there is reason to hope that randomized approaches can also provide strong results in the dynamic case.

\subsubsection*{\bf Developing Intuition.} A common theme in the design of randomized algorithms for contention resolution is to decompose the dynamic case into a sequence of static cases. Here, we describe one method by which this decomposition is accomplished (for examples, see~\cite{BenderKPY18,bender:contention,bender2020contention}).

When a packet $u$ arrives, it listens to the channel to determine whether it is already in use by other packets. Any packets that have arrived ahead of $u$ may still be attempting to succeed and, if so, they should also be emitting a ``busy signal'', which we describe momentarily. If $u$ detects this busy signal, then it listens until this signal ceases, at which point it can start attempting to transmit. In this way, there are often two groups of packets: (1) \defn{active} packets that are trying to transmit, while maintaining a busy signal, and (2) \defn{idle} packets that are waiting for their turn on the channel. Given this isolation, active packets can execute a contention-resolution algorithm for the static case in order to succeed; afterwards, any idle packets can become active and repeat this process.

\begin{figure}[t]
\centering
\includegraphics[scale=0.27]{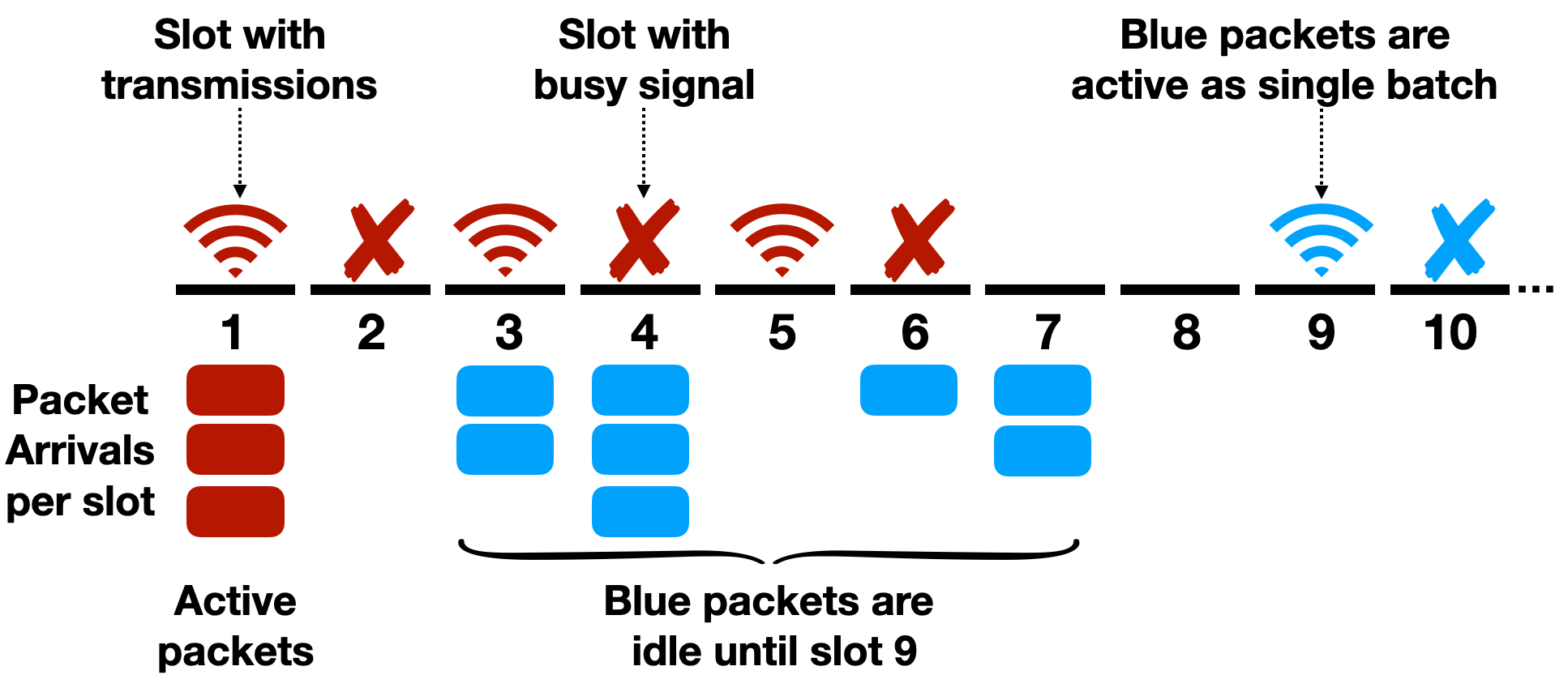}
\vspace{-5pt}\caption{An example of how a busy signal may be used to decompose the dynamic case into a series of static cases. Here, three packets (colored red) arrive in slot 1 and issue the busy signal in slots 2, 4, and 6. During this time, other packets (colored blue) arrive, but remain idle, since they detect the busy signal. Once all red packets succeed, slot 8 is empty, and the blue packets become active.}\label{fig:ran-intuition}
\end{figure}

The busy signal can take different forms. For example, given CD and a global clock,  active packets can emit a busy signal in each even-numbered slot; see Figure~\ref{fig:ran-intuition}. Any idle packet that hears this signal or a collision infers that active packets are present and keeps waiting. However, having all packets send in every second slot may not be desirable, since sending consumes energy~\cite{SINHA201714,feeney2001investigating,polastre:telos,wang2006realistic}.  To reduce this cost, packets may emit the busy signal probabilistically, which can greatly reduce the sending cost; however, there is now a probability of error in the situation where some packets may still be active, but fail to emit a busy signal. When this happens, any idle packets will become active prematurely and interfere with the already-active packets. 

This interference is mitigated in different ways in the literature. Bender et al.~\cite{BenderKPY18,bender:contention} assume CD and $O(\log n)$ control bits (recall Section~\ref{s:alg-design}) in each transmission on the shared channel. The busy signal occurs probabilistically every $O(1)$ slots, with each active packet sending an expected $O(1)$ times to maintain the signal. Interleaved with the busy signal are slots used to estimate the number of active packets, $m\leq n$, which allows them to execute \tstb (recall Section~\ref{sec:backoff-backon}) starting with a window of size $\Theta(m)$. Additionally, a leader is elected from the set of active packets to broadcast how many slots remain in the execution of \tstb; this serves as a busy signal during the execution of \tstb. The leader is frequently changed so that each packet is responsible for $O(1)$ transmissions in expectation. Newly-arrived idle packets need only listen for a handful of slots to receive this information, which allows them to (i) avoid becoming active prematurely and (ii) avoid any further listening until the current execution of \tstb is finished. While the above process can fail---for example, the estimate is poor, the busy signal fails, or active packets do not succeed in the execution of \tstb---this occurs with  probability at most polynomially small in the number of active packets,  which allows for $O(n)$ expected latency. Additionally, the expected number of channel accesses performed per packet is $O(\log\log^* n)$.  This low number of accesses can be important to modern networks where devices are battery-powered, given that listening and sending on the channel consume significant energy. 

{\revfontg  We note that while a busy signal has been used to achieve strong algorithmic results, it may be problematic in practice, since a constant fraction of the active slots are wasted on this synchronization.}

\subsubsection{Randomized $k$-Selection}
We extend our discussion above by continuing to survey the results that employ randomized algorithms for the dynamic $k$-selection problem.  In the context of randomized algorithms,  results focus the case where $k=n$. 
We note that the jamming-resistant contention-resolution results (presented next in Section \ref{sec:jamming}) clearly apply when there is no jamming, although we defer their presentation until later.




De Marco and Stachowiak~\cite{DeMarco:2017:ASC:3087801.3087831} explore randomized solutions with no-CD. When $n$ is known (or a known to within a constant multiplicative factor), they provide a non-adaptive algorithm with $O(n)$ latency, which is asymptotically optimal.  For the case where $n$ is unknown, a lower bound of $\Omega(n\log n/(\log\log n)^2)$ latency is shown to hold for any non-adaptive, randomized algorithm; thus, there is an asymptotic separation in latency between adaptive and non-adaptive randomized algorithms. Additionally, for $n$ unknown, the authors provide a strongly non-adaptive, non-acknowledgment-based algorithm with $O(n\log^2 n)$ latency; this latency can be improved to  $O(n\log^2 n/(\log\log n))$ with an acknowledgment-based algorithm, which comes close to matching their lower bound. Lastly,  when $n$ is unknown, the authors provide an adaptive algorithm that achieves $O(n)$ latency. {\revfontg Informally, this algorithm employs LE to break the dynamic problem into a sequence of static contention-resolution problems, as discussed earlier in this section.}

A more recent result by Bender et al. \cite{bender2020contention} shows that \whp constant throughput can be achieved by an adaptive algorithm with no-CD.  In contrast to the adaptive algorithm (for unknown $n$) of De Marco and Stachowiak~\cite{DeMarco:2017:ASC:3087801.3087831}, 
the result by Bender et al. \cite{bender2020contention} allows packets to terminate immediately upon success.

When CD is available, Bender et al.~\cite{bender:contention} achieve $O(n)$ latency using expected $O(\log\log^* n)$ channel accesses per packet; this is shown to be asymptotically optimal by Chang et al.~\cite{ChangJP19}. As discussed in Section \ref{sec:metrics}, this  low number of channel accesses translates is important to modern networks where devices are battery-powered, given that listening and sending on the channel consumes energy. {\revfontg We highlight that LE plays a key role in keeping energy costs low.}

In a multi-channel setting, with and without CD, Christodoulou et al.~\cite{christodoulou2018strategic} consider a game theoretic setting where \things may deviate from any prescribed protocol. The authors give protocols from which no \thing will unilaterally deviate and that provide probabilistic latency guarantees.

Finally, recent work by Bender et al. \cite{bender:coded} investigates a problem variant where at most an integer $\kappa\geq 1$ simultaneous successful transmissions are possible. This model is reminiscent of~\cite{hradovich2020contention} (in Section~\ref{sec:aqt-results}), and it captures newer radio technology that can decode a limited number of concurrent transmissions. Here, the notion of CD alters slightly: \things can distinguish between an empty slot and a slot with multiple transmissions, but they do not learn whether the number of transmissions in any slot exceeds or is at most $\kappa$ (unless they are one of the senders). Here, the authors show that a throughput of $1-o(1)$ is possible.

  
\subsubsection{The Wakeup Problem}\label{sec:wakeup-problem}

There is a known separation between static and dynamic wakeup. Specifically,  recall that for static wakeup with no-CD, Kushilevitz and Mansour \cite{kushilevitz1998omega} show a lower bound of $\Omega(\log n)$, and an asymptotically matching upper bound is achieved via a randomized algorithm by Bar-Yehuda et al. \cite{bar1992time}. However, for dynamic wakeup, Jurdzinski and Stachowiak \cite{jurdzinski2015cost} prove a lower bound of $\Omega(\log^2 n)$ in the case of no-CD.

Under no-CD, Jurdzi{\'n}ski and Stachowiak~\cite{jurdzinski2005probabilistic,jurdzinski2002probabilistic} present randomized algorithms with probability of error $\epsilon$ for solving the wakeup problem under different settings: where \things are labeled or unlabeled, where there is a global clock or not,  and where $n$ is known or unknown. We note that, while deterministic approaches rely on the use of labels, this is an example of labels being used in a randomized solution; in this case, the labels are used to approximate system size.  The authors are concerned with uniform algorithms (recall Section \ref{sec:static-wakeup}). {\revfontg  Perhaps unsurprisingly, aspects of these wakeup algorithms are reminiscent of exponential backoff.} One of the main findings is a near-exponential separation between the global and local clock settings when $n$ is unknown and there no labeling; the former can be solved with latency $O( \log(n)\log(1/\epsilon))$, while the latter requires $ \Omega(n/\log n)$ latency. In the cases where $n$ is known, or the \things are labeled (or both), the run time is $O(\log n \log(1/\epsilon))$. For uniform algorithms,  the authors also show a nearly-tight lower bound of $\Omega\left(\frac{\log(n) \log(1/\epsilon)}{\log\log(n) + \log\log(1/\epsilon)}\right)$ in the case where $n$ is known and the \things are labeled.  As stated earlier in Section \ref{sec:static-wakeup}, this bound improved by Farach-Colton et al. \cite{farach-colton:lower-bounds} to $\Omega(\log(n) \log(1/\epsilon))$, and then later shown to hold for all randomized algorithms by Newport \cite{newport2014radio,newport:radio-journal}.


Under the SINR model (recall Section \ref{sec:static-wakeup}), and in the absence of a global clock and assuming \things know a polynomial upper bound on the system size, Jurdzi{\'n}ski and Stachowiak \cite{jurdzinski2015cost} give a Monte Carlo algorithm that has \whp $O(\log^2 n/\log \log n)$ latency, and a Las Vegas algorithm with expected $O(\log^2 n/\log \log n)$ latency, which may be contrasted with the deterministic algorithm in \cite{jurdzinski2013distributed} of $O(\log^2 n)$ latency (recall Section \ref{sec:det-wakeup}).

For a single-hop radio network, Chlebus et al.~\cite{Chlebus:2016:SWM:2882263.2882514} consider $k$ devices becoming active over time,  with access to a known number of $b$ channels. The model used is no-CD and no global clock. Here, all stations may access all channels at once. When $k$ is known, Chlebus et al. present a randomized algorithm with error $\epsilon$ that has latency $O(k^{1/b}\ln(1/\epsilon))$.  

Chrobak, G\k{a}sieniec, and Kowalski~\cite{ChrobakGK07} consider the wakeup problem for $n$ devices in a multi-hop radio network with no global clock and no collision detection. Each node has a unique label from a range of size $O(n)$. The authors present a randomized algorithm with run time $O(D\log^2 n)$, where $D$ is the network diameter.

\medskip 

\subsubsection*{\bf Summary.} For the $k$-selection problem, a key message is that we can get asymptotically optimal $O(k)$ latency using adaptive algorithms.  For the wakeup problem, the situation is also significantly improved by randomization. Specifically, compared to the results surveyed in Section~\ref{sec:det-wakeup}, we see that randomization often allows for $O(\texttt{polylog}(n))$ latency in contrast to $\omega(n)$ latency in the single-hop scenario; however, in the local clock setting without knowledge of key parameters, the latency remains high. That said, in comparison to the results from the AQT and deterministic approaches, it is clear that randomization can offer significant improvements. 

Another aspect of randomized approaches to the $k$-selection problem  is the focus on the number of access attempts, which translates to energy consumption.  The $O(\log\log^*n)$ complexity achieved by Bender et al.~\cite{bender:contention} exploits the ability to send and receive messages of $\Theta(\log n)$ bits. Under the more standard ternary feedback model, the best results achieve a polylogarithmic in $n$ number of access attempts.

Finally, we note that while many of the randomized results for $k$-selection achieve asymptotically-optimal latency, they are fragile in the sense that even a small amount of channel interference can cause the corresponding algorithm to fail. For example, the result of Bender et al.~\cite{bender:contention} estimates the number of packets contending for channel access; even a single slot with interference can cause a gross overestimate. Another instance is a result by De Marco and Stachowiak \cite{DeMarco:2017:ASC:3087801.3087831}, which depends on coordination messages issued by a leader; such messages can be disrupted by interference.  The algorithm by  Bender et al.~\cite{bender2020contention} is also unable to maintain asymptotically-optimal latency in the face of jamming.  Thus, robustness to malicious interference presents a challenge, and it is the subject we explore next.





\section{Jamming-Resistant Contention Resolution}\label{sec:jamming}

\begin{itemize}[leftmargin=3.4cm]
\item[]{\it ``I bought the jammer from a local electrical store and asked the police if it was OK and they said it was - it's great as it has stopped the problem, but some of the local shops aren't happy.''}\\ 
\mbox{\hspace{2.4cm} -- Father Michele Madonna of Santa Maria di Montesanto Church~\cite{priest}}
\end{itemize}

A jamming attack is characterized by deliberate collisions on the multiple access channel. By deviating from a prescribed protocol, a jamming device (\defn{jammer}) can launch a denial-of-service attack, preventing other devices from being able to communicate~\cite{DBLP:reference/algo/RichaS16}. There have been several well-publicized instances of jamming~\cite{FCC:humphreys,marriott,priest,nicholl}, behavior that may   become more common given the trend towards wireless personal devices. For example, in reference to the quote above, Father Madonna installed a store-bought jamming device in his church to prevent cellphone usage by attendees from disrupting his services. However, this device also interfered with nearby wireless devices, with shop owners complaining that credit card payment systems were affected.

There are a number of papers addressing applied security considerations with respect to jamming (e.g., \cite{bayraktaroglu:performance,xu:feasibility,aschenbruck:simulative,mpitziopoulos:survey,VADLAMANI201676}); for example, spread spectrum techniques~\cite{spread:foiling,navda:using}, rerouting defenses~\cite{wood2003jam,han2018hierarchical}, and specialized antenna technology~\cite{zheng2018design,mcmilin2015gps}. From a theory perspective, several fundamental distributed computing challenges---such as broadcast~\cite{koo2,gilbert:near,gilbert:making,chen:broadcasting}, leader election~\cite{gilbert:malicious,gilbert2009malicious}, gossip~\cite{gilbert:interference,dolev:gossiping}, node discovery~\cite{meier:speed}, and simple point-to-point communication~\cite{king:conflict,DBLP:journals/dc/KingPSY18}---can be solved on a shared channel despite jamming (see  \cite{young:overcoming,Bender:2015:RA:2818936.2818949}). Here, we focus on prior contention-resolution results  that offer provable performance guarantees. 

\subsubsection*{\bf Developing Intuition.} A jamming adversary poses at least two significant new challenges over the dynamic-arrivals model considered in Section~\ref{sec:adversarial-dynamic}. First, it may not be obvious that non-trivial latency guarantees are possible; for example, if jamming occurs in every slot in perpetuity, then no \things can succeed. Therefore, we must either constrain the jamming power of the adversary (e.g., by assuming a bounded adversary), or alter our measure of throughput to address this case (e.g., via notions of competitiveness and implicit throughput). Second, the algorithms surveyed so far that rely on channel feedback to adjust contention---for example, by estimating $n$ (such as in~\cite{BenderKPY18,DBLP:journals/jacm/GreenbergFL87}) or by exchanging messages between \things (such as in \cite{BenderKPY18,DeMarco:2017:ASC:3087801.3087831}---are vulnerable, since jamming allows the adversary to cause overestimates of $n$ and prevent communication. 

To address the first challenge, a common approach is to optimize performance in non-jammed slots, as done by Chang et al.\cite{ChangJP19} and Bender et al.\cite{bender:how}. Assuming that at least one \thing is present in the system, a slot that is not jammed, but where no \thing succeeds, is said to be \defn{wasted}.
Guaranteeing that only a small constant fraction of slots are wasted allows for meaningful results despite an adversary that jams aggressively.

Addressing the second challenge of adjusting the contention requires care, and this is demonstrated by the elegant algorithm of Chang et al. \cite{ChangJP19}. \Things follow some simple rules for updating the sending probability in each slot: increasing if there is silence, decreasing if there is noise, and leaving it unchanged in the case of a success. The adversary may lower the sending probabilities of \things by jamming (i.e., adding noise), thus depressing contention. However, the update rule allows for contention to quickly recover once the jamming ceases (although, it may resume again later), and few slots are wasted as a result. Notably, this design allows the algorithm to achieve roughly $1/e$ throughput in the non-jammed slots, which rivals that of slotted ALOHA, despite the more-challenging adversarial setting.

\subsubsection*{\bf Survey of the Literature.} 

Broadly speaking, the jamming adversaries treated in the literature on contention resolution can be categorized as \defn{bounded} or \defn{unbounded}. The former has constraints that limit the amount of jamming; typically, this is specified by a jamming rate or by a fraction of the slots that can be jammed within any contiguous set of slots. In contrast, as the name suggests, an unbounded adversary may jam arbitrarily.

An early series of results against a bounded adversary was initiated by Awerbuch {et al.} \cite{awerbuch:jamming,DBLP:journals/talg/AwerbuchRSSZ14}, who consider the following model. There are $n$ \things equipped with CD, and the network is saturated (recall Section~\ref{sec:general-monotonic-backoff}). For some $T \in \mathbb{N}$ and constant $0<\epsilon<1$, there is an adversary that can jam at most $(1-\epsilon)w$ of the slots in any consecutive $w\geq T$ slots. This adversary is referred to as  {\boldmath{$(T, 1 - \epsilon)$}}{\defn{-bounded}}, and the value of $\epsilon$ is unknown to \things. 

 Such bounded jamming may be well-motivated in practice, where aggressive jamming can  reveal that an attack is occurring, as well as the location of the malicious device(s) (for examples, see~\cite{liu2009localizing,liu2013error,xu:feasibility}), or correspond to a limited power supply. The adversary is also \defn{adaptive}, meaning that the adversary knows the full history of the actions performed by \things for all the previous slots, and may use this information to inform its jamming strategy in the current slot; however, the adversary does not know what actions \things are taking in the current slot. When the adversary jams a slot, all listening \things hear jamming; this is referred to as a \defn{1-uniform} adversary.

Awerbuch~{et al.}~\cite{awerbuch:jamming,DBLP:journals/talg/AwerbuchRSSZ14} define an algorithm to be \defn{constant competitive} if, for any sufficiently large number of slots, \things succeed in a constant fraction of the non-jammed steps. The authors provide a randomized algorithm that is \whp constant competitive when all \things  know a common parameter, {\boldmath{$\gamma$}}$= O(1/(\log T + \log\log n))$, and the algorithm is run for a sufficiently large  number of slots.  Roughly speaking, in a slot, the algorithm dictates that each \thing $u$ sends a message with probability $p_u$. If $u$ does not send, it listens and updates $p_u$ via a multiplicative factor which can increase or decrease, depending on whether or not the slot is idle.  Notably, this algorithm design shares common ground with the later work by Chang et al.~\cite{ChangJP19} discussed above.

Subsequently, this line of work has been extended to several challenging settings, showing constant competitiveness so long as $\gamma$ is known to all the \things, and the algorithm is executed for a sufficiently large number of slots.  Richa~{et al.}~\cite{richa:jamming2} consider a multi-hop network modeled as a  unit disk graph (UDG). Informally, under the UDG model, each \thing can broadcast within a disk centered about its location, and two \things may communicate if their respective disks intersect (see Schmid and Wattenhofer~\cite{schmid2006algorithmic} for discussion of the UDG model). First, Richa~{et al.}~\cite{richa:jamming2,DBLP:journals/dc/RichaSSZ13} consider the setting where the $(T, 1 - \epsilon)$-bounded adversary is 1-uniform,  and the UDG is connected or each \thing has at least $2/\epsilon$ neighbors. The authors provide an algorithm, named \textsc{Jade}, that is \whp constant competitive. Second, Richa~{et al.}~\cite{richa:jamming2} prove that, if the graph is not connected, or the adversary is $2$-uniform---for any slot, the adversary selects any partition of the $n$ \things and has the members of one partition hear jamming, while the others do not---and some \things have $o(1/\epsilon)$ neighbors, then  \textsc{Jade} can fail to be constant competitive. Against a \defn{reactive} jamming adversary---who first learns whether the channel is idle in the current slot and then chooses whether to jam---Richa~{et al.}~\cite{richa:jamming3,DBLP:journals/ton/RichaSSZ13} show that constant competitiveness is possible. 

Under the SINR channel model (recall Section~\ref{sec:static-wakeup}), 
Ogierman~et al.~\cite{ogierman:competitive,ogierman2018sade} address jamming in a more-general network topology than prior results on competitiveness. The authors provide a multiplicative weights update algorithm that is $2^{-O(1/\epsilon)^{2/(\alpha-2)}}$\hspace{-5pt}-competitive, where $\alpha>2$ is a parameter of the SINR model known as the path-loss exponent, which corresponds to the degree of signal attenuation. This competitiveness holds if the algorithm is executed for a sufficiently large number of slots against a $1$-uniform adversary and the transmission range contains at least one node, or at least $2/\epsilon$ nodes within the transmission range of every node. 

In the case where multiple networks inhabit the same terrain, bandwidth may be scarce and inter-network communication may not be possible. In this setting,  Richa~{et al.}~\cite{richa:jamming4} present a multiplicative weights update algorithm that allows each of the $K$ networks to obtain $\Omega(\epsilon^2\min\{\epsilon, 1/\mbox{poly}(K)\})$-fraction of the non-jammed slots.

A different approach is taken by Bender et al.~\cite{bender:how,BenderFGY19}, who address an adversary that is not constrained in its jamming. A packet starts in an ``inactive'' state by listening to the channel, and if a sufficient number of empty slots are heard, the packet becomes ``active'' in the sense of making transmission attempts. This ``throttling'' mechanism gives time for contention to decrease, thus allowing active packets to succeed. An active packet $u$ will make transmission attempts, but in contrast to updating based on feedback in each slot, $u$ monotonically reduces its sending probability  as roughly $1/s_u$, where $s_u$ is the number of slots since $u$ became active. Packet $u$ also listens (when not transmitting), and if a sufficient number of empty slots are heard, it resets $s_u=1$. Here, they show expected constant throughput, with each packet making an expected  number of transmission attempts that is polylogarithmic in $n+J$, where $J$ is the total number of jammed slots during which there exists at least one packet. Here, again, it is worth highlighting the related result by Chang et al. \cite{ChangJP19} improves on this number of transmission attempts.

Recently, Bender et al. \cite{bender:fully} showed how to achieve $\Omega(1)$ implicit throughput  (recall Section \ref{sec:metrics}) with $O(\poly{\log n})$ channel accesses (both sending and listening), despite an adaptive jamming adversary. This result applies in the channel setting with CD (i.e., ternary channel feedback), but with no control bits allowed, unlike the energy-efficient result of Bender et al. \cite{BenderKPY18}. The algorithm, named \textsc{Low-Sensing Backoff},  is based on multiplicative weights updates, and in that regard, it shares  common ground with some prior results (such as \cite{ChangJP19,richa:jamming2,richa:jamming3,richa:jamming4, awerbuch:jamming,ogierman2018sade}). However, in comparison to these results,  \textsc{Low-Sensing Backoff} greatly limits the frequency at which packets listen to the channel in order to update, while the lack of control bits greatly limits the amount of information they receive in each update. Analogous results are given for the cases where the number of packets is infinite, and where the adversary is reactive.


The AQT setting (recall Section~\ref{sec:aqt-results}) has been extended to incorporate jamming. Anantharamu et al.~\cite{anantharamu2019packet,anantharamu2011medium} extends a leaky-bucket adversary   to {\boldmath${(\lambda,\rho, T)}$}\defn{-bounded} leaky bucket adversary who can also jam $\lambda t+b$ slots in any window of consecutive $t$ slots; here, $\lambda$ is the jamming rate, while $b$ is the burstiness parameter. Deterministic algorithms are presented, with and without CD, giving a range of bounds on queue size and latency parameterized by the injection rate, burstiness, and the {\it known} number of stations $n$, each of which has a unique ID in the range $[0, ..., n-1]$.

In a related setting, but with a single sender and receiver, Anta et al.~\cite{DBLP:journals/tcs/AntaGKZ17} consider an adversary  whose jamming capability is parameterized by a rate, $\rho \in [0, 1]$,  and the length of the largest burst of jammed packets, $\sigma \geq 1$; both parameters are known to the algorithm.  Each packet has a fixed-length header, along with a payload whose size may be adjusted by the sender. The authors derive  a lower bound on the optimal rate of transmission for payload information (i.e., goodput), and also develop an adaptive deterministic algorithm that is optimal.

Jurdzi\'nski et al.~\cite{DBLP:conf/waoa/JurdzinskiKL14} also consider a single sender and receiver, but in contrast to a parameterized adversary, they consider an {\it unbounded} adversary and a setting where packets of {\it different} sizes may arrive over time. If the adversary jams at any time during which a packet is transmitted, the transmission fails and the sender learns of this immediately. 
The authors derive results on the ratio between the throughput achieved by the algorithm and the throughput achieved by the best scheduling under identical arrival and error patterns; this is referred to as relative throughput. The authors give a deterministic online algorithm that achieves optimal relative throughput. 

An impossibility result is given by Bender et al.~\cite{bender2020contention}, who show that with no-CD, no algorithm can guarantee a packet success in time linear in the system size with better than constant probability in the presence of jamming; this result holds even with a non-adaptive adversary. Subsequently, Chen et al. \cite{DBLP:conf/podc/ChenJZ21} give an algorithm in the no-CD case that achieves a trade-off between the throughput and the amount of jamming tolerated. In the challenging case where a constant fraction of slots are jammed,  $O(1/\log n)$ throughput is shown, which aligns with the above lower bound.

\subsubsection*{\bf Summary.} Perhaps unsurprisingly, randomization along with CD can provide constant throughput (in non-jammed slots) despite an unbounded adversary; even without CD, jamming can be tolerated, although throughput worsens slightly to $\Theta(1/\log n)$. When the adversary is bounded, a range of results exist. Under a saturated model, randomized results demonstrate constant throughput for a variety of network settings against a bounded adversary, so long as some weak information about the adversary's power and $n$ is known. In contrast, deterministic results in the AQT setting are more nuanced, being parameterized by aspects of jamming rate and burstiness, and generally more pessimistic.

Most results that tolerate jamming exhibit a sharp trade-off between the amount of listening devices perform and throughput achieved. On one hand,  if each \thing listens to the channel in every slot (except when sending), then  constant throughput (in expectation or \whp in non-jammed slots) is possible~\cite{bender:how,BenderFGY19,ChangJP19,awerbuch:jamming,richa:jamming2,richa:jamming3}. Note that such results are not energy efficient, given that listening is roughly as costly as sending in practice (recall Section~\ref{sec:metrics}).  On the other hand, if devices never listen to the channel, only sub-constant throughput is possible, and the best result is $O(1/\log n)$ throughput \whp~\cite{DBLP:conf/podc/ChenJZ21}. The recent result in \cite{bender:fully} demonstrates that jamming-resistant contention resolution is possible with constant throughput and a polylogarithmic number of channel accesses (both sending and listening) without using control bits. However, it is unclear whether better energy-efficiency bounds are possible in the presence of jamming, perhaps under models with additional feedback permitted on the channel.


\section{Conclusion}

We have surveyed the literature on adversarial contention resolution by delineating results based on key model assumptions, metrics, and algorithm-design choices. The problem of contention resolution remains an active topic of research, and here we briefly discuss some potential directions for future work.

Machine learning (ML) may have an important role to play in the design of contention-resolution algorithms; {\revfontg certainly, the use of ML is being explored in the development of future wireless standards, such as 6G \cite{kato:ten}}. As discussed in Section~\ref{sec:wakeup-problem}, Gilbert et al.~\cite{gilbert2021contention}  demonstrate the impact of system-size predictions, which might be offered by a black-box ML component. A natural question is whether there are other useful predictions that can give improved performance, {\revfontg or at least performance that is parameterized by the accuracy of the predictions. This has been done recently for denial-of-service attacks in the traditional client-server model \cite{chakraborty:bankrupting}.} Can predictions about whether a slot will be noisy (or jammed) yield improved robustness? 

{\revfontg In a similar vein, incorporating capabilities enabled by new (but existing) wireless technology can yield novel results (e.g., coded networks \cite{bender:coded}) and deserves further exploration. For instance, the 5G New Radio (NR) ability to dynamically change the width of the frequency band being used can help mitigate congestion or interference on the channel, and allow for power savings (e.g., \cite{abinader:impact}). Defining a clean abstract model for these capabilities and studying their effects on fundamental problems like contention resolution could be a valuable research focus.}

In many wireless systems, collisions have significant cost. For example, in WiFi networks, the entire packet must be sent before the sender learns of success or failure (due to a collision). For instance, in mobile settings, failure to transmit a packet may mean incurring a significant delay before the sender is next within range of the intended receiver. Therefore, modeling and minimizing collision cost presents some interesting open problems.

{\revfontg Mobility adds a layer of complexity to wireless communication because \things may be constantly changing position, affecting signal strength and interference. This is an aspect that has not received attention in the literature, and integrating mobility into contention resolution might open the door to interesting new problems. Additional costs due to distance traveled or energy expended to overcome attenuation are aspects that may be worthwhile considering.}

{\revfontg Adopting different channel models can yield distinct upper and lower bounds compared to the classical model, as we have seen in the case of the SINR model (e.g., \cite{fineman:contention,jurdzinski2015cost}). The power-sensitive model 
explored by De Marco and Kowalski \cite{DBLP:conf/icdcs/MarcoK10} and by Tsybakov \cite{tsybakov:multiplicity} is another example of an  interesting channel model. In the latter case, exploring via experiment the degree to which this model holds in practice may be interesting, as it seems to impart a non-trivial amount of power (e.g., we can count the number of users in $O(1)$ slots rather than $\poly{\log n}$ slots). Even coarser-grained feedback ---rather than an exact value---for the number of senders could be useful. 
For example, in the classical channel, it is hard to tell if noise comes from many users competing for access or from jamming; however, (even rough) estimates could help clarify this.
}

Another dimension to consider is traffic priority. Consider a packet that contains voice-over-IP data, and another packet holding old data being backed up on a remote server. Which packet should get access to the shared channel first? Intuitively, the first packet is time sensitive and should have priority; if it is delayed too long, the packet contents are useless. This prioritization may become increasingly important to  wireless networks, as a combination of limited bandwidth and a growing number of devices forces a differentiated treatment of traffic. While this aspect has been explored by Agrawal et al. \cite{DBLP:conf/spaa/AgrawalBFGY20}, there are several outstanding questions regarding robustness to adversarial jamming, heterogeneous packet sizes, and potential trade-offs with the amount of delay packets may suffer as a result.

Energy efficiency is likely to remain an important consideration in wireless systems.  The exact limits of trading off between robustness to worst-case noise, energy efficiency, and latency still remain unknown. Can we obtain $o(\log n)$ number of channel-access attempts per packet despite jamming, while retaining asymptotically optimal latency?  {\revfontg Related to this aspect,  in some wireless standards, transmission power can be adjusted, allowing devices to compensate for signal attenuation. This capability might be useful in overcoming jamming attacks, where good \things incur higher energy costs to transmit at higher power in order to overcome malicious noise. Is it possible to design algorithms in the spirit of \cite{ChangJP19,BenderFGY19}, where the sending costs are grow very slowly with the jamming costs? Alternatively, can the latency be asymptotically improved over existing jamming-resistant results?} 

\subsubsection*{\bf Acknowledgements.} The authors are grateful to the anonymous reviewers for their feedback. 





\clearpage

\section{Electronic Supplement}\label{sec:electronic-supplement}

Our electronic supplement contains our tables, each of which serves as a quick reference for many of the results presented in major sections of our survey. For each such section, in addition to a brief summary, we select aspects from Section~\ref{sec:prelim} that give context for the corresponding result. For results where an algorithm is not explicitly specified as adaptive, full-sensing, or acknowledgement-based (or where some ambiguity may be present), we mark the corresponding "algorithm type" with a dash in our tables.


\subsection*{Table for Section~\ref{sec:backoff-algorithms}: The Static Case}

\begin{table}[h!] 
\begin{center}
{
\begin{tabular}{  |P{4.65cm}|P{4.65cm}|  }
\hline
\rowcolor{LightCyan} {\bf Algorithm} &  {\bf Makespan}  \\
\hline
\B & $\Theta(n\log n)$   \\
\hline
\LB & $\Theta\left( \frac{n\log n}{\log\log n} \right)$ \\
\hline
\LLB & $\Theta\left( \frac{n\log\log n}{\log\log\log n} \right)$ \\
\hline
\FB & $\Theta\left( n\log\log n\right)$ \\
\hline
\PB & $\Theta\left(\frac{n}{\log n}\right)^{1+(1/c)}$ \\
\hline
\STB  & $\Theta\left(n\right)$ \\
\hline
\TSTB  & $\Theta\left(n\right)$ \\
\hline
\textsc{One-fail Adaptive} & $O(n)$  \\
\hline
Algorithm of~\cite{Anta2010} &   $O(n + \log^2(1/\epsilon))$  \\
\hline
\end{tabular}
}\vspace{4pt}\caption{Makespan results under the static model with $n$ packets for randomized algorithm. For \PB, $c$ is a constant strictly greater than $1$. Makespan results hold with high probability in $n$; in the case of \cite{Anta2010}, the makespan holds with probability $1-O(\epsilon)$, where $\epsilon \leq 1/(n+1)$.}\label{table:makespan}\vspace{-8pt}
\end{center}
\end{table}


\begin{table}[h!] 
\begin{center}
\renewcommand\arraystretch{1.3}
\begin{tabular}{ |P{2.5cm}|P{2cm}| >{\raggedright}p{7.5cm} | }
\hline
\rowcolor{LightCyan} {\bf Paper} &  {\bf Channel Feedback} &  {\bf Summary of Result} \tabularnewline
\hline
Willard~\cite{willard:loglog} 
&   CD    & Lower bound on expected latency is at least $\log \log k$, and upper bound is $\log\log k + O(1)$, for $2 \leq k \leq n$.
\tabularnewline 
\hline
 Fineman et al.~\cite{fineman:contention2}  
& CD &  Latency of $O\left(\frac{\log n}{\log \mathcal{C}} + (\log\log n)\right)$ for $k = 2$, and $O\left( \frac{\log n}{\log \mathcal{C}} + (\log\log n)(\log\log\log n)\right)$ latency for $k\leq n$.   \tabularnewline
 \hline
Kushilevitz and Mansour~\cite{kushilevitz1998omega,kushilevitz1993omega}&  
No-CD& When global clock present, expected latency is $O(D\log n)$ for non-uniform algorithm, where $D$ is the diameter of the radio network. \tabularnewline
\hline
Bar-Yehuda et al. \cite{bar1992time}& No-CD & Expected $O(\log n)$ makespan, and \whp $O(\log^2 n)$ makespan. \tabularnewline
\hline
Fineman et al.~\cite{fineman:contention}& --- &  Under the SINR channel model, \whp makespan is $\O(\log n + \log R)$,  where $R$ is typically $O(\poly n)$. \tabularnewline
\hline 
\end{tabular}
\vspace{4pt}\caption{Classification of the literature on static wakeup  (Section~\ref{sec:static-wakeup}).}
    \end{center}
\end{table}


\clearpage
\subsection*{Table for Section~\ref{sec:aqt-results}: AQT and Contention Resolution }


\begin{table}[h!] 
\begin{center}
{
\begin{tabular}{ |P{1.8cm}|P{1.7cm}|P{1.1cm}|P{2.1cm} |>{\raggedright}p{6.3cm}|}
\hline
\rowcolor{LightCyan} {\bf Paper} & {\bf Algorithm Type} & {\bf Channel Feedback} & {\bf  Adversary Type } &  {\bf Summary of Result} \tabularnewline
\hline
 
Bender et al.~\cite{bender:adversarial} & --- 
& No-CD  & Leaky bucket & \beb is stable for $\rho = d/\log n$, where $d>0$ is a constant, and unstable for $\rho = \Omega(\log\log n/\log n)$. \llb is stable for $\rho=d'/( (\log n)\log\log n)$, when $d'>0$ is a constant, and instability for $\rho = \Omega(1/\log n)$.
\tabularnewline
\hline

\multirow{3}*{} Chlebus et al.~\cite{chlebus2009maximum,chlebus2007stability}\footnote{Possibility results hold with no-CD, whereas impossibility result holds regardless CD or no-CD.}
 & Ack-based & No-CD  & Window, where $\rho = 1$  & For $n=2$ station, stability impossible for burstiness $b\geq2$. \tabularnewline \cline{2-5}
 & Full sensing & No-CD &  Window, where $\rho = 1$  & For $n=2$ stations, fair latency possible. For  $n=3$ stations, stability impossible for burstiness $b\geq2$. \tabularnewline \cline{2-5}
 & --- 
 & No-CD  &  Window, where $\rho = 1$  & For $n=3$ stations, fair latency possible. For $n \geq 4$ stations, stability possible. For $n \geq 4$ stations, stability and fair latency impossible. \tabularnewline\cline{2-5}
 & Ack-based & No-CD  &  Leaky bucket, where $\rho = 1$  & Stable and fair latency impossible, except $n=1$ station case.  \tabularnewline \cline{2-5}
 & Full sensing & No-CD  &   Leaky bucket, where $\rho = 1$   & For $n\geq2$ stations, stability impossible for burstiness $b\geq2$.\tabularnewline \cline{2-5}
 & --- 
 & No-CD  &   Leaky bucket, where $\rho = 1$   & For $n\geq2$ stations, stability possible and $O(n^2+b)$ packets queued at any given time, but both stability and fair latency impossible.
  \tabularnewline 
\hline

\multirow{3}*{} Anantharamu and Chlebus~\cite{anantharamu2015broadcasting}
 & Non adaptive \& activation-based&  CD  & $1$-activating  & For $\rho < 1/3$
 and $b\geq 3$, the latency is bounded by $(3b-3)/(1-3\rho)$, and the maximum queue size is $(3b-3)/2$.  \tabularnewline \cline{2-5}
 & Non adaptive \& full sensing &  CD  &  $1$-activating &  Stable and bounded latency of $2b+4$ and there are at most $b+O(1)$ packets queued, for  $\rho \leq 3/8$. 
 \tabularnewline \cline{2-5}
  & Adaptive \& activation-based &  CD and No-CD  &  $1$-activating &  Stable and bounded latency of at most $4b-4$ and at most $2b-3$ packets in the queued, for $\rho \leq 1/2$. \tabularnewline \cline{2-5}
  & Adaptive  \& full sensing & CD and No-CD &  $1$-activating & Stability and bounded latency not possible. \tabularnewline
\hline

\end{tabular}
}\vspace{4pt}\caption{Classification of the literature on AQT with respect to contention resolution (Section~\ref{sec:aqt-results}).}
\end{center}
\end{table}\clearpage

\begin{table}[t!] 
\begin{center}
{
\begin{tabular}{ |P{2cm}|P{1.7cm}|P{1.4cm}|P{2cm} |>{\raggedright}p{6.3cm}|}
\hline
\rowcolor{LightCyan} {\bf Paper} & {\bf Algorithm Type} & {\bf Channel Feedback} & {\bf  Adversary Type } &  {\bf Summary of Result} \tabularnewline
\hline

\multirow{4}*{}  Chlebus et al.~\cite{chlebus2006adversarial,chlebus:adversarial}
 & Full sensing & CD  & Window, where $\rho < 1$  & Fair latency and stability possible for $\rho = O(1/\log n)$.\tabularnewline \cline{2-5}
 
 & Full sensing & No-CD  & Window, where $\rho < 1$   &  Fair latency and universal stability possible for $\rho = O(1/c\log^2 n)$, where $c$ is sufficiently large. Explicit algorithm proposed for $\rho = O(1/\poly{\log n})$.\tabularnewline \cline{2-5}
 
 & Ack-based  & CD &  Window, where $\rho < 1$   & Stability impossible for $\rho > 3/(1+\log n)$, for $n\geq 4$. \tabularnewline 
\hline

\multirow{3}*{} Anantharamu et al.~\cite{anantharamu2009adversarial,anantharamu2017adversarial}
 & Full sensing \& ack-based & No-CD  & Window  & Stability with lower bound latency $\Omega(w \max\{1, \log_w n\})$ for $w\leq n$ and $\Omega(w+n)$ for $w>n$.  \tabularnewline \cline{2-5}
 & Full sensing \& adaptive & No-CD  &  Window &  Stability achieved with upper bound latency $O(\min\{n+w, w\log n\})$.  \tabularnewline \cline{2-5}
  & Full sensing \& non-adaptive & No-CD  &  Window & Algorithms with $O(n + w)$ queue size  and $O(nw)$ latency. \tabularnewline 
\hline
 
\multirow{3}*{} Hradovich et al.~\cite{hradovich2020contention}
 & Adaptive & No-CD  & Leaky bucket  & Stability for $\rho \leq 1$ with $O(n^2 + b)$ queued packets.  \tabularnewline \cline{2-5}
 & Full sensing and ack-based &  CD &  Leaky bucket & Stability for $\rho \geq 1-1/n$ with throughput $\Theta(k/n(\log^2 n))$ on a $k$-restrained channel.  \tabularnewline \cline{2-5}
  & Ack-based  & No-CD  &  Leaky bucket & Throughput for deterministic, acknowledgement-based algorithms with a global clock cannot be better than $\min(k/n,1/(3\log n))$ on a $k$-restrained channel.\tabularnewline 
\hline

Bienkowski et al.~\cite{bienkowski:distributed,BienkowskiJKK12} & --- & No-CD & Unrestricted & Total load is  additive $O(n^2)$ larger and maximum queue size is $n$ times bigger with additive $O(n)$ larger than any offline algorithm in distributed queue setting. 
\tabularnewline
\hline

\multirow{3}*{} Garncarek et al.~\cite{garncarek2018local}
 & Adaptive & No-CD  & Leaky bucket  & With $O(n\log (L_{max}))$ bits (small memory), stability is possible with bounded queue size $L_{max}$, for $\rho < 1$ and $b \geq 0$. \tabularnewline \cline{2-5}
 & Non-adaptive &  No-CD & Leaky bucket & Without memory, stability is possible for $\rho = O(1/\log n)$.\tabularnewline \cline{2-5}
  & Non-adaptive  & No-CD  & Leaky bucket & Without memory and with ultra strong selectors, stability is possible for $\rho \leq d/\log^2 n$ for some constant  $d>0$.\tabularnewline 
\hline

\multirow{3}*{}
Garncarek et al.~\cite{garncarek2019stable}
 & --- & No-CD & Leaky bucket  & Stability for $\rho = \Omega(1/\log n)$ with $O(1)$ memory and $b$ is unknown. \tabularnewline \cline{2-5}
 & ---  &  No-CD & Leaky bucket & Stable for $\rho = O(1)$ with no memory and bounded $b$. 
 \tabularnewline 
\hline

\end{tabular}
}\vspace{4pt}\caption{Classification of the literature on AQT with respect to contention resolution (Section~\ref{sec:aqt-results}).}
\end{center}
\end{table}


\clearpage

\subsection*{Table for Section~\ref{sec:deterministic}: Deterministic Approaches}

\begin{table}[h!] 
\begin{center}
\renewcommand\arraystretch{1.3}
\begin{tabular}{ |P{2cm}|P{1.9cm}|P{1.4cm}|P{1.0cm}|P{1cm}| >{\raggedright}p{6cm} | }
\hline
\rowcolor{LightCyan} {\bf Paper} & {\bf Algorithm Type} & {\bf Channel Feedback} &  {\bf Global Clock}  &  {\bf {\boldmath{$k$}} Known} & {\bf Summary of Result} \tabularnewline
\hline
\rowcolor{gray!20!}
    \multicolumn{6}{|c|}{\textbf{Results for \boldmath{$k$}-Selection Problem}} \\
     \hline
Greenberg and Winograd~\cite{greenberg1985lower}  & Adaptive  & CD    &  Yes   & Yes    &  Lower bound on latency of $\Omega(k\log n/\log k)$. \tabularnewline
 \hline
De Marco and Kowalski~\cite{doi:10.1137/140982763}  & Non-adaptive   & No-CD    & Yes   & No   & Dynamic channel guarantees $O(k\log n\log\log n)$ latency. \tabularnewline

\hline
\multirow{3}*{} De Marco et al. \cite{DBLP:conf/icdcs/MarcoKS19}
& Ack-based \& non-adaptive & No-CD  & No  & Yes & Upper bound latency is $O(k\log k \log n)$. \tabularnewline \cline{2-6}
& Ack-based \& non-adaptive & No-CD  & No  & No & Upper bound latency is $O(k^2 \log n/\log k)$ and lower bound latency is $\Omega(\min\{k^2 \log kn,n\}+k^2/ \log k)$. \tabularnewline \cline{2-6}
&  Non-ack-based \& non-adaptive  & No-CD & No & Yes  & Upper bound latency is  $O(k^2 \log n)$ and lower bound latency is $\Omega(\min \{k^2\log kn,n\})$. \tabularnewline \cline{2-6}
&  Non-ack-based or ack-based \& non-adaptive  & No-CD & No & No  & Lower bound latency is $\Omega(k^2/\log k)$. \tabularnewline 
\hline


\rowcolor{gray!20!}
    \multicolumn{6}{|c|}{\textbf{Results for Wakeup Problem}} \\
\hline
     
De Marco, Pellegrini, and Sburlati~ \cite{DBLP:journals/dam/MarcoPS07,DBLP:conf/ictcs/MarcoPS05}& Non-adaptive  & No-CD   &  No   &   No  &   For single-hop radio network with upper bound latency $O(n^3\log^3n)$, when n is unknown.\tabularnewline

\hline
Chlebus et al.~\cite{Chlebus:2016:SWM:2882263.2882514}& Non-adaptive  &  No-CD  &  No   &  No   & For single-hop radio network with upper bound latency $O(k\log^{1/b}k \log n)$. For  $b=\Omega(\log\log n)$ channels, faster algorithm attained with latency $O((k/b) \log n \log(b\log n))$ and lower bound latency $((k/b)\log(n/k))$.  \tabularnewline

\hline

Chlebus et al.~\cite{ChlebusGKR05}& Non-adaptive &  No-CD   &  Yes   &  N/A   & Multi-hop ad-hoc network with latency $O(n\log^2 n)$. Explicit algorithm provides latency $O(n\Delta \poly{\log n})$, where $\Delta$ is the maximum in-degree neighbors. \tabularnewline

\hline

Jurdzi{\'n}ski and Stachowiak \cite{jurdzinski2015cost}& Adaptive  &   CD  &  Yes/No   &  ---   &  SINR channel model,  show $\Omega(n)$ latency without a global clock,; show $\Omega(\log^2 n)$ latency with a global clock. \tabularnewline
\hline

\end{tabular}
\vspace{4pt}\caption{Classification of the literature on deterministic approaches (Section~\ref{sec:deterministic}).}
    \end{center}
\end{table}

\clearpage

\begin{table}[t!] 
\begin{center}
\renewcommand\arraystretch{1.3}
\begin{tabular}{ |P{2cm}|P{1.9cm}|P{1.4cm}|P{1.0cm}|P{1cm}| >{\raggedright}p{6cm} | }
\hline
\rowcolor{LightCyan} {\bf Paper} & {\bf Algorithm Type} & {\bf Channel Feedback} &  {\bf Global Clock}  &  {\bf {\boldmath{$k$}} Known} & {\bf Summary of Result} \tabularnewline
\hline

\rowcolor{gray!20!}
    \multicolumn{6}{|c|}{\textbf{Results for Wakeup Problem}} \\
\hline

Chrobak, G\k{a}sieniec, and Kowalski~\cite{ChrobakGK07}& Non-adaptive  &  No-CD  &   No  &  N/A    &  Multi-hop ad-hoc network with upper bound latency is $O(n^{5/3}\log n)$.   \tabularnewline

\hline
Chlebus et al.~\cite{ChlebusK04}& Non-adaptive  &   No-CD  &  No   &  N/A   & Multi-hop ad-hoc network with (a non-constructive) upper bound on latency of $O(n^{3/2}\log n)$, and a construction with $O(k^2 \poly{\log n})$ latency.
\tabularnewline

\hline

\multirow{4}*{} De Marco and Kowalski~\cite{DEMARCO20171,de2013contention}
 & Non-adaptive& No-CD & No  & No & Upper bound latency $O(k \log(n/k) + 1)$ when the time that the first \thing becomes active is known; achieve $O(k \log(n) \log\log(n))$  when the time that the first \thing becomes active is unknown. \tabularnewline \cline{2-6}
 & Non-adaptive & No-CD & No  & Yes & Upper bound latency $O(k \log(n/k) + 1)$ when the time that the first \thing becomes active is unknown.\tabularnewline \cline{2-6}
& --- & --- & --- & --- & A lower bound of $\Omega((k \log(n/k) + 1))$  latency. \tabularnewline
\hline
\end{tabular}
\vspace{4pt}\caption{Continuation: Classification of the literature on deterministic approaches (Section~\ref{sec:deterministic}).}
    \end{center}
\end{table}

\clearpage

\subsection*{Table for Section~\ref{sec:random-stream}: Randomized Approaches}

\begin{table}[h!] 
\begin{center}
{
\begin{tabular}{ |P{2cm}|P{1.4cm}|P{2.0cm}| >{\raggedright}p{6.8cm} | }
\hline
\rowcolor{LightCyan} {\bf Paper} & {\bf Channel Feedback} &  {\bf System Size  Knowledge } &  {\bf Summary of Result} \tabularnewline
\hline
    \rowcolor{gray!20!}
    \multicolumn{4}{|c|}{\textbf{Results for \boldmath{$k$-Selection Problem}}} \\
     \hline
Bender et al.~\cite{BenderKPY18,bender:contention}  & CD & No  & Expected $O(\log(\log^* n))$ channel accesses (sending and listening) per packet and expected $O(n)$ latency. \tabularnewline
 \hline
\multirow{3}*{} De Marco et al.~\cite{DeMarco:2017:ASC:3087801.3087831}
 & No-CD  & No  &  
Lower bound of $\Omega(n\log n/(\log\log n)^2)$ latency  for any non-adaptive, randomized algorithm; a strongly non-adaptive, non-acknowledgment-based algorithm with $O(n\log^2 n)$ latency; this latency can be improved to  $O(n\log n/(\log\log n)^2)$ with an acknowledgment-based algorithm; and an adaptive algorithm that achieves $O(n)$ latency.\tabularnewline 
\hline
Bender et al.~\cite{bender2020contention} & No-CD & No & The $n_t$ packets achieve $\Omega(1)$ throughput in first $t$ steps with a total of  $O(n_t\log^{2}n_t)$ channel transmission attempts.\tabularnewline
 \hline
Bender et al.~\cite{BenderFGY19,bender:how}& CD & No & Expected constant throughput and expected $O(\log^2 (n+D))$ send attempts per packet, where $D$ denotes the unavailable slots due to adversary. For the case of infinite packet arrivals, expected $O(\log^2 (\eta + D))$ send attempts per packet, where $\eta$ represents the maximum number of packets ever in the system concurrently.\tabularnewline
\hline
 Bender et al.~\cite{bender:coded}  & ---  & No  & In a radio coded network, for any window with $w \geq 16\kappa^2$ slots, suppose there are at most $1 - \left(\frac{5}{\ln \kappa}\right)w$ packet arrivals, where $\kappa$ is the maximum number of packets that can successfully be simultaneously sent. Then, this result guarantees \whp in $w$ that: 
 (1) at any given time, the number of packets in the system is at most $2w$, and (2) each packet succeeds within $O(\sqrt{\kappa}w\ln^3w)$ time slots.\tabularnewline
\hline
\end{tabular}
}\vspace{4pt}\caption{Classification of the literature on randomized approaches (Section~\ref{sec:random-stream}).}
\end{center}
\end{table}

\begin{table}[t!] 
\begin{center}
{
\begin{tabular}{ |P{2.3cm}|P{1.4cm}|P{1.7cm}| >{\raggedright}p{6.5cm} | }
\hline
\rowcolor{LightCyan} {\bf Paper}  & {\bf Channel Feedback} &  {\bf  System Size Knowledge } &  {\bf \hspace{2cm}Summary of Result} \tabularnewline
\hline
\rowcolor{gray!20!}
    \multicolumn{4}{|c|}{\textbf{Results for Wakeup Problem}} \\
     \hline
     Jurdzi{\'n}ski and Stachowiak \cite{jurdzinski2015cost}
     &No-CD&$ O(\poly n)$ known
     & Lower bound of $\Omega(\log^2 n)$ latency for any randomized algorithm using classical (i.e., non-SINR) channel model.  For the SINR channel, a Monte Carlo algorithm that has \whp $O(\log^2 n/\log \log n)$ latency, and a Las Vegas algorithm with expected $O(\log^2 n/\log \log n)$ latency. \tabularnewline
\hline
\multirow{2}*{} Jurdzi{\'n}ski and Stachowiak \cite{jurdzinski2005probabilistic,jurdzinski2002probabilistic}
& No-CD & ---  &  
When $n$ is unknown and there no labeling, access to a 
global clock yields $O( \log(n)\log(1/\epsilon))$ latency, while with local clock the latency is  $ \Omega(n/\log n)$.
In the cases where $n$ is known or the \things are labeled (or both), the run time is $O(\log n \log(1/\epsilon))$. \tabularnewline \cline{2-4}
 & No-CD & Yes  & For uniform algorithms,  with $n$ known and things labeled, there is a lower bound of $\Omega\left(\frac{\log(n) \log(1/\epsilon)}{\log\log(n) + \log\log(1/\epsilon)}\right)$ latency.\tabularnewline 
\hline
Chrobak, G\k{a}sieniec, and Kowalski~\cite{ChrobakGK07}
&No-CD&Yes& Multi-hop radio network with no global clock and packets are labeled, latency is $O(D \log^2 n)$, where $D$ is the diameter of the network. \tabularnewline
\hline
Chlebus et al.~\cite{Chlebus:2016:SWM:2882263.2882514}
& No-CD & Yes & In single hop radio network with no global clock, latency is $O(n^{1/b}\ln(1/\epsilon))$, where $b$ is number of channel and $\epsilon$ is the error. \tabularnewline
\hline
\multirow{2}*{} Gilbert et al.~\cite{gilbert2021contention}
 & CD & --- & Expected number of rounds required for $X$ (random variable) participants is lower bounded by $\frac{H(c(X))}{2}-O(\log\log\log\log n)$ and upper bound by $O(H^2(c(X)))$, where $H(c(X))$ represents the entropy of the condensed probability distribution over the possible network sizes.\tabularnewline\cline{2-4}
 & No-CD & --- & Expected number of rounds required for $X$ (random variable) participants is lower bounded by $\Omega\left(\frac{2^{H(c(X))}}{\log\log n}\right)$ and upper bounded by $O\left(2^{2H(c(X))}\right)$. \tabularnewline 
\hline

\end{tabular}
}\vspace{4pt}\caption{Continuation: Classification of the literature on randomized approaches (Section~\ref{sec:random-stream}).}
\end{center}
\end{table}

   
\clearpage

\subsection*{Table for Section~\ref{sec:jamming}: Jamming-Resistant Contention Resolution}


\begin{table}[h!] 
\begin{center}
{
\begin{tabular}{ |P{2cm}|P{1.6cm}|P{1.4cm}|P{1.9cm}|P{1.9cm}| >{\raggedright}p{5.3cm} | }
\hline
\rowcolor{LightCyan} {\bf Paper} & {\bf Jamming Type} & {\bf Channel Feedback} &  {\bf Deterministic or  Randomized}  &  {\bf System-Size Knowledge} & {\bf Summary of Result} \tabularnewline
\hline

Awerbuch et al. \cite{awerbuch:jamming,DBLP:journals/talg/AwerbuchRSSZ14} & $(T, 1-\epsilon)$-bounded, adaptive &  CD & Randomized  &  Upper bound & Constant-competitive algorithm for single-hop networks.\tabularnewline

\hline

Richa et al.
\cite{richa:jamming2,DBLP:journals/dc/RichaSSZ13} &  $(T, 1-\epsilon)$-bounded, adaptive  & CD & Randomized  & Upper bound  & In multi-hop connected unit disk graph (UDG) network, constant competitiveness achieved if executed for sufficiently large number of slots  when the adversary is $1$-uniform and at least $2/\epsilon$ nodes within the transmission range. Constant competitiveness not achieved if UDG is disconnected or there are nodes with $o(1/\epsilon)$ nodes within transmission range.\tabularnewline 

\hline

Richa et al. 
\cite{richa:jamming3,DBLP:journals/ton/RichaSSZ13}& $(T, 1-\epsilon)$-bounded, reactive &  CD & Randomized &  Upper bound & Achieves $e^{-\Theta(1/\epsilon^2)}$-competitiveness and near-fairness if executed for sufficiently large number of slots.\tabularnewline

\hline
Richa et al. 
\cite{richa:jamming4} & $(T, 1-\epsilon)$-bounded, adaptive  &  CD  & Randomized &  Upper bound & In $K$ co-existing network can obtain $\Omega(\epsilon^2\min\{\epsilon, 1/\mbox{poly}(K)\})$-fraction of the non-jammed slots.
\tabularnewline

\hline

Ogierman et al.
\cite{ogierman:competitive,ogierman2018sade}& $(T, 1-\epsilon)$-bounded, adaptive &  CD & Randomized & Upper bound &  In SINR network, $2^{-O(1/\epsilon)^{2/\alpha-2}}$-competitiveness achieved if executed for a sufficiently large number of slots against a $1$-uniform adversary and transmission range contains at least one node, or at least $2/\epsilon$ nodes within the transmission range of every node. No MAC protocol can achieve any throughput against a $(B,T)$-bounded adversary, where $B$ is the energy budget.
\tabularnewline 

\hline

Chang et al. 
\cite{ChangJP19}& Unbounded, adaptive &  CD & Randomized  & No & Throughput  of $1/e-O(\epsilon)$ in non-jammed slots, where $\epsilon$ is step size.\tabularnewline 

\hline Anta et al. 
\cite{DBLP:journals/tcs/AntaGKZ17} & 
$(\sigma,\rho)$-bounded, adaptive & --- & Deterministic &  --- & Setting with a single sender and receiver. The main result is showing a goodput of roughly $(1 - \sqrt{f/T})^2$ under cases where packet length is fixed, when $f$ is the number of errors introduced by the adversary, and  $T>f$ is length of time interval. \tabularnewline

\hline
\end{tabular}
}\vspace{4pt}\caption{Classification of the literature on jamming-resistant contention resolution (Section~\ref{sec:jamming}).}
\end{center}
\end{table}
\clearpage

\begin{table}[t!] 
\begin{center}
{
\begin{tabular}{ |P{1.9cm}|P{1.6cm}|P{1.4cm}|P{1.9cm}|P{1.8cm}| >{\raggedright}p{5cm} | }
\hline
\rowcolor{LightCyan} {\bf Paper} & {\bf Jamming Type} & {\bf Channel Feedback} &  {\bf Deterministic or  Randomized}  &  {\bf System-Size Knowledge} & {\bf Summary of Result} \tabularnewline

\hline
Jurdzinski et al. 
\cite{DBLP:conf/waoa/JurdzinskiKL14}& Unbounded, adaptive  &  --- & Deterministic &  Exact & Setting with a single sender and receiver, and with packets of arbitrary lengths are sent. The main result is an algorithm that achieves optimal relative throughput. \tabularnewline

\hline

Anantharamu et al. 
\cite{anantharamu2019packet,anantharamu2011medium}& $(\rho,\lambda,b)$-leaky bucket-bounded  & No-CD  & Deterministic & Exact &  Several deterministic algorithms are designed and analyzed, with and without CD, giving a range of bounds on queue size and latency parameterized by the injection rate, burstiness, and the {\it known} number of stations $n$.\tabularnewline

\hline
Chlebus et al.
\cite{Chlebus:2016:SWM:2882263.2882514}&Random & No-CD & Deterministic &  Upper bound & Two algorithms in multi-channel with $b$ channels achieves. With high probability in $n$, the first has latency $O\left(\log^{-1}\left(\frac{1}{p}\right)k\log n\log^{1/b}k\right)$. The second has latency $O\left(\log^{-1}\left(\frac{1}{p}\right)\left( \frac{k} {b}\right)\log n\log(b\log n)\right)$ for $b>\log(128b\log n)$,  where $p>0$ is a known probability of jamming in any slot.\tabularnewline

\hline
Bender et al.~\cite{bender2020contention}& Unbounded & No-CD & Randomized & No & Constant throughput without CD, where \things terminate upon success. Also, it is shown that no algorithm can guarantee a packet success in time linear in the system size with better than constant probability in the presence of jamming.\tabularnewline 
\hline
Chen et al.~\cite{DBLP:conf/podc/ChenJZ21} & Unbounded &No-CD & Deterministic &  No & A major result is that throughput of $O(1/\log n)$ \whp is possible in the no-CD case. \tabularnewline 
\hline

Bender et al.~\cite{BenderFGY19,bender:how}& Unbounded & CD & Randomized & No & Expected constant throughput and expected $O(\log^2 (n+J))$ send attempts per packet, where $J$ is number of jammed slots. For the case of infinite packet arrivals, expected $O(\log^2 (\eta + J))$ send attempts per packet, where $\eta$ represents the maximum number of packets ever in the system concurrently.\tabularnewline
\hline

Bender et al. \cite{bender:fully} & Unbounded & CD & Randomized & No & With high probability in $n+J$, where $J$ is the number of jammed slots,  constant (implicit) throughput is possible with $\poly{\log n}$ sending and listening attempts per packet.
\tabularnewline 

\hline
\end{tabular}
}\vspace{4pt}\caption{Continuation: Classification of the literature on jamming-resistant contention resolution (Section~\ref{sec:jamming}).}
\end{center}
\end{table}


\end{document}